\documentclass[prb,twocolumn,showpacs]{revtex4}
\usepackage[latin1]{inputenc}
\usepackage{graphicx}
\usepackage{amssymb}
\usepackage{amsmath}
\usepackage{xspace}
\usepackage{dcolumn}
\usepackage{bm}

\newcommand{\ie}[0]{i.e.\@\xspace}
\newcommand{\eg}[0]{e.g.\@\xspace}
\newcommand{\Z}[0]{\mathcal{Z}}
\newcommand{\D}[0]{\mathcal{D}}
\newcommand{\omb}[0]{\bar{\omega}}
\newcommand{\etal}[0]{{\it et al.}\@\xspace}
\newcommand{\tr}[0]{\text{Tr}\,}
\newcommand{\om}[0]{\omega}
\newcommand{\si}[0]{\sigma}
\newcommand{\las}[0]{\langle}
\newcommand{\ras}[0]{\rangle}
\newcommand{\la}[0]{\left\las}
\newcommand{\ra}[0]{\right\ras}
\newcommand{\ket}[1]{\left|#1\ra}  
\newcommand{\bra}[1]{\la#1\right|} 
\newcommand{\sket}[1]{|#1\ras}  
\newcommand{\sbra}[1]{\las#1|} 
\newcommand{\dtau}{\Delta\tau}
\newcommand{\rmd}{\mathrm{d}}
\newcommand{\rmi}{\mathrm{i}}
\newcommand{\Rs}{\mathbb{R}}
\newcommand{\op}{\hat{p}}
\newcommand{\ox}{\hat{x}}
\newcommand{\on}{\hat{n}}
\newcommand{\wb}{w_\text{b}}
\newcommand{\wf}{w_\text{f}}
\newcommand{\UU}{{\cal U}}

\begin{document}


\title{Quantum Monte Carlo and variational approaches to the Holstein model}

\author{Martin Hohenadler}\email{hohenadler@itp.tu-graz.ac.at}
\author{Hans Gerd Evertz}
\author{Wolfgang \surname{von der Linden}}
\affiliation{%
  Institute for Theoretical Physics, Graz University of Technology,
  Petersgasse 16, A-8010 Graz, Austria} 

\begin{abstract}
  Based on the canonical Lang-Firsov transformation of the Hamiltonian we
  develop a very efficient quantum Monte Carlo algorithm for the Holstein
  model with one electron. Separation of the fermionic degrees of freedom by
  a reweighting of the probability distribution leads to a dramatic reduction
  in computational effort.  A principal component representation of the
  phonon degrees of freedom allows to sample completely uncorrelated phonon
  configurations.  The combination of these elements enables us to perform
  efficient simulations for a wide range of temperature, phonon frequency and
  electron-phonon coupling on clusters large enough to avoid finite-size
  effects.  The algorithm is tested in one dimension and the data are
  compared with exact-diagonalization results and with existing work.
  Moreover, the ideas presented here can also be applied to the
  many-electron case.  In the one-electron case considered here,
  the physics of the Holstein model can be described by a simple variational
  approach.
\end{abstract}


\maketitle

\section{\label{sec:introduction}Introduction}

Quantum Monte Carlo (QMC) simulations for models with electron-phonon
coupling are often limited in both system size and accessible parameter range
by long autocorrelation times and large statistical errors. This makes it
very difficult to study realistic models for, \eg, the high temperature
superconductors or the manganites which exhibit colossal magnetoresistance.
In both classes of materials electron-phonon interactions play an important
role.\cite{BYMdLBi92,David_AiP} Although classical treatments of phonons have
been quite successful in certain situations,\cite{MiMuSh96II,DaYuMobook} \eg,
at high-enough temperatures, quantum effects are expected to be
relevant.\cite{David_AiP} Consequently, it is highly desirable to develop a
new, more efficient method to treat the phonon degrees of freedom quantum
mechanically. The Holstein molecular-crystal model constitutes one of the
simplest models for coupled electron-phonon systems, and therefore serves as
an ideal testing ground for new approaches.  Moreover, despite enormous
theoretical efforts, even the physics of the Holstein model is still not
completely understood.

The extensive use of QMC methods to study many-body problems is based on the
fact that they can give quasiexact results (\ie, exact apart from statistical
errors which can be made arbitrarily small, in principle). Over the years,
several different QMC methods have been developed to study systems with
electron-phonon coupling, such as the Holstein,\cite{Ho59a} the Fr\"ohlich,
\cite{FrPeZi50} or the Su-Schrieffer-Heeger (SSH) model.\cite{SuShHe79}  A
very general QMC method for coupled fermion-boson models was developed by
Blankenbecler \etal\cite{BlScSu81} and Scalapino and Sugar.\cite{ScSu81} It
is based on an analytic integration over the fermion degrees of freedom and a
MC simulation of the resulting boson model.  The simulation is performed
using the grand-canonical ensemble and requires the evaluation of a fermion
determinant involving a computation time which is proportional to the cube of
the system size. Moreover, the method in its original form becomes unstable
at low temperatures. While the simulations of
Refs.~\onlinecite{BlScSu81} and~\onlinecite{ScSu81} were restricted to one dimension, Levine
and Su\cite{LeSu90,LeSu91} and, using a stabilized version of the same
algorithm, Niyaz \etal\cite{NiGuScFo93} studied charge-density-wave formation
and superconductivity in the two-dimensional Holstein model.  A numerically
faster method is the world-line algorithm developed by Hirsch
\etal\cite{HiScSuBl81,HiSuScBl82} based on a special breakup of the
Hamiltonian and a fixed number of fermions.  This results in configuration
weights which are simple to evaluate allowing for much bigger system sizes.
In the course of the simulation, both fermions and bosons are sampled
simultaneously.  The latter method has been successfully applied to the
Holstein polaron problem and to the half-filled SSH and Holstein
model.\cite{HiScSuBl81,HiSuScBl82,HiFr82,HiFr83I,HiFr83II}  However the
world-line algorithm is restricted to models in one spatial dimension or to
the single-electron case in any dimension by the minus-sign problem.\cite{wvl1992}
Scalettar \etal\cite{ScBiSc89} applied a rather complicated so-called hybrid
molecular dynamics algorithm to the two-dimensional Holstein model near
half filling. This work was extended to the low-temperature regime by Noack
\etal\cite{NoScSc91} Finally, Marsiglio\cite{Mars90} developed a
low-temperature QMC method to study the same model, also at half filling.  De
Raedt and Lagendijk\cite{dRLa82,dRLa83,dRLa84} and Kornilovitch\cite{Ko97}
used an alternative approach based on Feynman's path-integral
method,\cite{Feynman55} where the boson degrees of freedom are integrated out
analytically and the resulting
fermionic model is simulated using QMC. Although the method is limited to one
electron or two electrons of opposite spin\cite{dRLa86}
by the sign problem, it allows efficient simulations in one, two, and
three dimensions even for small phonon frequencies near the adiabatic limit,
and has also been used to investigate the Holstein model with dispersive
phonons.\cite{dRLa84} Also using Feynman's path integral, Kornilovitch and
Pike\cite{KoPi97} developed a QMC method which exploits the conservation of
the total quasimomentum of the system and allows the calculation of
dynamical properties such as, \eg, the polaron band structure. Although the
method is not restricted to a certain model or dimensionality of the system,
it suffers from large statistical errors. Moreover, it is limited to the case
of a single fermion at very low temperature, and exhibits a sign problem for
nonzero total quasimomentum even in one dimension. Prokof'ev and
Svistunov\cite{PrSv98} and Mischenko \etal\cite{MiPrSaSv00} used QMC to
directly sample the zero-temperature one-electron Green function of the
Fr\"ohlich polaron in imaginary time. The method allows calculations for an
infinite system in three dimensions, but requires a convergent series for the
electron propagator. While all but the last method mentioned so far make use
of the Trotter-Suzuki approximation,\cite{wvl1992}
Kornilovitch\cite{Ko98,Ko99} developed a continuous-time algorithm that works
in any dimension and allows calculations on infinite systems. It gives
directly dynamical quantities such as the polaron spectrum and effective mass
with very high accuracy. However, similar to previous work,\cite{KoPi97} it
is restricted to the one-electron limit at very low temperatures. Moreover,
calculations for small phonon frequencies and/or weak electron-phonon
coupling are difficult and a sign problem appears for nonzero total
quasimomentum. The projector QMC method\cite{wvl1992} in combination with a
local updating of the phonon degrees of freedom has been used by Berger
\etal\cite{BeVavdL95} to investigate the Holstein-Hubbard model at various
band fillings, and Green function QMC simulations for the half-filled
Holstein model of spinless fermions have been performed by McKenzie
\etal\cite{KeHaMu96}  Finally, the stochastic series expansion MC technique
has been applied recently to an extended, one-dimensional Hubbard model with
an electron-phonon interaction of the SSH type.\cite{SeSaCa03}  In contrast
to other work, the phonons are treated in second quantization. Although the
method allows simulations on large lattices in one dimension, it relies on an
upper limit for the number of phonons at each site which makes it difficult
to study the regime of small phonon frequencies and/or strong coupling.

In addition to the specific shortcomings of each method such as, \eg, the
restriction to a single fermion, or to one spatial dimension, or to zero
temperature, all previous simulations of the Holstein model were limited to
some extent by autocorrelations. If the phonon degrees of freedom are not
integrated out analytically, these correlations predominantly come from the
free harmonic-oscillator dynamics, especially in the adiabatic regime of
small phonon frequency.  This often leads to an enormous computational effort
even for rather small systems and intermediate temperatures.

In this paper we first present a simple variational approach using a
generalized form of the Lang-Firsov transformation which, in the one-electron
case considered here, gives surprisingly good results and yields valuable
insight into the mechanism of polaron formation.  The full Hamiltonian
resulting from the standard version of the canonical Lang-Firsov
transformation is then used as the starting point for a QMC method which
is free of any uncontrolled approximations.  Due to the fact that the
Lang-Firsov transformation contains the crucial electronic influences on the
phonons, the Monte Carlo simulation for the phonon degrees of freedom can be
based only on the purely phononic part of the transformed Hamiltonian. The
electronic contributions can then be allowed for by reweighting of the
probability distribution, corresponding to an exact treatment of the fermion
degrees of freedom.  This enables us to completely ignore the electronic
weights in the updating process, and thereby dramatically reduce the
computational effort. Finally, we introduce a principal component
representation of the phonon coordinates, which allows exact sampling of the
phonons and avoids all autocorrelations.

The paper is organized as follows. We briefly review the Holstein model in
Sec.~\ref{sec:holstein}. In Sec.~\ref{sec:lang-firs-transf} we discuss
the aforementioned transformations of the Hamiltonian. Section~\ref{sec:VPA} is
devoted to the variational polaron approach, while the QMC method for the
Lang-Firsov transformed model is presented in Sec.~\ref{sec:LFQMC}.
Section~\ref{sec:reweighting} describes the reweighting method, and in
Sec.~\ref{sec:princ-comp-repr} the representation of the phonons in principal
components is introduced. Results obtained with the given methods are presented
in Secs.~\ref{sec:resVPA} and~\ref{sec:resQMC}. Finally,
Sec.~\ref{sec:summary} contains our conclusions.

\section{\label{sec:holstein}The Holstein model}

The Holstein model has been introduced in the 1950's,\cite{Ho59a} and
describes a system of tight-binding conduction electrons coupled to a
dispersionless phonon mode. If we express the phonon operators in terms of
their natural units, the Hamiltonian takes the form
\begin{eqnarray}\label{eq:holstein}\nonumber
  H &=& K+P+I\,,\\\nonumber
  K &=&-t\sum_{\las ij\ras \si}c^\dag_{i\si} c_{j\si}\,,\\\nonumber
  P &=& P_\text{p}+P_\text{x}
  =\frac{\om}{2}\sum_i\left( \op_i^2 + \ox_i^2 \right)\,,\\
  I &=&-\alpha\sum_i \on_i \ox_i\,.
\end{eqnarray}
Here $c^\dag_{i\sigma}$ ($c_{i\sigma}$) creates (annihilates) an electron of
spin $\si$ at lattice site $i$, $\ox_i$ and $\op_i$ denote the displacement
and momentum of a harmonic oscillator at site $i$, and $\on_i=\sum_\si
\on_{i\si}$ with $\on_{i\si}=c^\dag_{i\si} c_{i\si}$. The last term $I$
describes the local coupling of the dispersionless Einstein phonons to the
electron density $\on_i$.  In the first term, the symbol $\las ij\ras$
denotes a summation over all nearest-neighbor hopping pairs $(i,j)$ and
$(j,i)$.  The parameters of the model are the hopping integral $t$, the
phonon energy $\om$, and the electron-phonon coupling constant $\alpha$. We
introduce the commonly used dimensionless coupling constant $\lambda =
\alpha^2/(\om W)$, where $W=4td$ is the bare bandwidth in $d$ dimensions.  We
also define the dimensionless phonon frequency $\omb=\om/t$ and express all
energies in units of $t$. Thus the model depends on two independent
parameters, $\omb$ and $\lambda$. Throughout this paper periodic boundary
conditions in real space are assumed.

The Holstein model has been investigated intensively in the past, using a
large variety of methods. Due to the large amount of literature
available we restrict the discussion to the case of a single electron in
an otherwise empty lattice with which this paper is concerned. The latter is
generally known as the Holstein polaron problem
and still constitutes a complicated many-body problem. Weak-coupling
perturbation theory has been found to be accurate only for very small
coupling strengths $\lambda$ when
the phonon frequency is low.\cite{Marsiglio95} In the strong-coupling regime,
an adiabatic small-polaron approximation\cite{Ho59a,KaRa94} has been found to
work extremely well for small values of $\omb$ (Ref.~\onlinecite{AlKaRa94}),
while a perturbation theory based on the Lang-Firsov
transformation\cite{LangFirsov} gives accurate results for $\omb\gg1$
(Refs.~\onlinecite{Marsiglio95} and~\onlinecite{AlKaRa94}). Discrepancies remain, however, in
the regime of intermediate coupling and phonon frequency.\cite{Ma90}  To
bridge this gap, a lot of numerical work has been done using exact
diagonalization (ED) methods, density-matrix renormalization-group (DMRG)
studies, QMC methods and variational methods. ED is limited in the accessible
parameter range, since it requires a truncation of the Hilbert space
associated with the phonon degrees of freedom. With increasing
electron-phonon coupling strength, for example, more and more phonon states
have to be included to obtain converged
results,\cite{RaTh92,Marsiglio93,AlKaRa94,Marsiglio95,WeRoFe96,WeFe97,Robin97,CaStGr97,CaCiGr98}
which makes it difficult to study clusters of reasonable size in the strong
or even intermediate coupling regime, especially for small phonon
frequencies. At this point calculations based on DMRG set in, which use an
optimized phonon basis to reduce the size of the Hilbert
space.\cite{JeWh98,JeZhWh99,ZhJeWh99,WeFeWeBi00} Another possible approach
are the so-called cluster methods which exploit exact information on small
clusters to obtain approximate results for infinite systems.\cite{Stephan,HoAivdL03}
Moreover, a number of variational methods have been developed which give very accurate
results over a wide range of
parameters.\cite{BoTrBa99,KuTrBo,RoBrLi99I,RoBrLi99II,CaFiIa01,FeLoWe00,Ba02,Ba03,AcCuNoRo01}
As discussed in Sec.~\ref{sec:introduction}, various QMC methods have been developed for
the Holstein model. The polaron problem considered here has been investigated
by Hirsch \etal,\cite{HiScSuBl81,HiSuScBl82} De Raedt and
Lagendijk,\cite{dRLa82,dRLa83,dRLa84} Kornilovitch,\cite{Ko97} Kornilovitch
and Pike,\cite{KoPi97} Kornilovitch,\cite{Ko98,Ko99} and Mishchenko
\etal\cite{MiPrSaSv00} Finally, the Holstein polaron has also been studied in the
infinite-dimensional limit using dynamical mean-field
theory.\cite{CiPaFrFe97}

For the one-dimensional case, on which we will focus here, the general
picture emerging from these investigations is as follows (see, \eg,
Ref.~\onlinecite{ZhJeWh99}). Starting from the noninteracting system
($\lambda=0$) the electron is gradually dressed with a coherent multi-phonon
cloud as the coupling increases. For $\lambda<1$ and $\lambda/\omb<0.5$ the
resulting quasiparticle remains in a Bloch-like state, with the phonon cloud
giving rise to an increased effective mass. In the strong-coupling regime
($\lambda >1$ and $\lambda/\omb>0.5$) the electron becomes self-trapped by
the induced lattice distortion and this object---trapped electron plus
distortion---is usually called a small polaron.  The transition from weak to
strong coupling is continuous,\cite{Loe88} and the term ``large polaron'' is
often used to describe an electron dressed with a phonon cloud extending over
more than one lattice site. The polaronic effects become more dominant as the
phonon frequency approaches the adiabatic limit $\omb\rightarrow 0$. In QMC
simulations, small values of $\omb$ introduce two very different time scales
for the evolution of electrons and phonons, respectively. This gives rise to
the problems mentioned above and, in fact, many QMC simulations have been
done only for $\omb\gtrsim1$.

\section{\label{sec:lang-firs-transf} Extended Lang-Firsov transformation}

The canonical Lang-Firsov transformation\cite{LangFirsov} has been used
extensively to study the polaron problem. A well-known approximation due to
Holstein\cite{Ho59a} consists of replacing the transformed hopping term by
its expectation value with respect to a zero-phonon state, thus neglecting
phonon emission and absorption during the hopping process. This
approximation, which we shall call the Holstein-Lang-Firsov (HLF)
approximation, was found to give reliable results only in the strong-coupling
and/or nonadiabatic limit $\lambda,\omb\gg1$
(Refs.~\onlinecite{ZhJeWh99},~\onlinecite{WeRoFe96} and~\onlinecite{WeFe97}). More refined approaches based
on strong-coupling perturbation theory provide an accurate description of the
Holstein polaron over a large range of parameters.\cite{Marsiglio95,AlKaRa94}
In the limit $\lambda=\infty$, the hopping term in
Hamiltonian~(\ref{eq:holstein}) can be neglected, and the Lang-Firsov
transformation allows an exact solution of the resulting single-site
problem.\cite{Ma90} The transformation has also been used in combination with
numerical methods.\cite{dMeRa97,Robin97,FeLoWe00} However we are not aware of
any QMC simulation based on the transformed model.

We define the unitary operator
\begin{equation}\label{eq:U}
  U=e^{\nu}\,,\quad
  \nu=\rmi\sum_{ij}\gamma_{ij} \on_i \op_j\,,
\end{equation}
where $i$ and $j$ run over lattice sites, and with real parameters
$\gamma_{ij}$. $U$ as defined in Eq.~(\ref{eq:U}) has the form of a
translation operator. Given an electron at lattice site $i$, it corresponds
to a displacement of the harmonic oscillators at all sites $j$,
$j=1,\dots,N$, by $\gamma_{ij}$. Hence the transformation describes a
nonlocal phonon cloud surrounding an electron. This corresponds to the
well-known concept of a large polaron, which extends over more than one
lattice site. Using the transformation $\tilde{\hat{O}}= U\,\hat{O}\,U^\dag$ we find
for the transformed operators
\begin{eqnarray}\label{eq:trop}\nonumber
  \tilde{\ox}_i &=& \ox_i + \sum_j \gamma_{ij} \on_j\quad,\quad
  \tilde{\op}_i = {\op}_i\\
  \tilde{c}_{i\si}^\dag &=& c_{i\si}^\dag\, e^{ \rmi\sum_j\gamma_{ij}\op_j}\quad,\quad
  \tilde{c}_{i\si}       =  c_{i\si}     \, e^{-\rmi\sum_j\gamma_{ij}\op_j}\,.
\end{eqnarray}
Inserting these results into Eq.~(\ref{eq:holstein}), the transformed
Hamiltonian becomes
\begin{eqnarray}\label{eq:htilde}\nonumber
  \tilde{H} &=&
  \tilde{K}+ P+\tilde{I}_\text{ep}+\tilde{I}_{ee}\\\nonumber
  \tilde{K} &=& -t\sum_{\las ij\ras\si}
  e^{\rmi\sum_l(\gamma_{il}-\gamma_{jl})\op_l}\;c^\dag_{i\si}c_{j\si}\\\nonumber
  \tilde{I}_\text{ep} &=& \sum_{ij} \on_j
  \ox_i(\om\gamma_{ij}-\alpha\delta_{ij})\\
  \tilde{I}_\text{ee} &=& \sum_{ij} \on_i \on_j\left(
  \frac{\om}{2}\sum_l\gamma_{lj}\gamma_{li}-\alpha\gamma_{ij}\right)\,.
\end{eqnarray}
Here the term $\tilde{I}_\text{ep}$ describes the coupling between electrons
and phonons, while $\tilde{I}_\text{ee}$ represents an effective
electron-electron interaction. The Hamiltonian~(\ref{eq:htilde}) will be the
starting point for the variational polaron approach presented in the following
section.

A more suitable approach for QMC simulations, however, is given by requiring
that the electron-phonon terms cancel.  Then $\gamma_{ij} =
\gamma\delta_{ij}$ with $\gamma=\sqrt{\lambda W/\om}$ and we obtain the
standard Lang-Firsov transformation with the transformation operator
\begin{equation}\label{eq:LFop}
U_0=e^{\nu_0}\quad,\quad \nu_0 = \rmi\gamma\sum_i \on_i \op_i\,.
\end{equation}
In contrast to the extended polaron cloud, defined by Eq.~(\ref{eq:U}), now
only the oscillator at the site of the electron is affected. The transformed
Hamiltonian reads
\begin{eqnarray}\label{eq:LFHamiltonian}\nonumber
  \tilde{H} &=& \tilde{K}+P+Q\,,\\\nonumber
  \tilde{K} &=&-t\sum_{\las ij\ras\si}c^\dag_{i\si}c_{j\si}
  e^{\rmi\gamma(\op_i-\op_j)}\,,\\
  Q &=&-\frac{1}{2}\gamma^2\om\sum_i \on_i^2\,.
\end{eqnarray}
In the HLF or small-polaron approximation, the ground state of the
transformed Hamiltonian is approximated by leaving all phonons in the ground
state.  It has been shown\cite{ZhJeWh99} that the small-polaron wave function
becomes exact in the strong-coupling, nonadiabatic limit, and agrees
qualitatively with the exact results also in the intermediate coupling
regime.  As discussed by Zhang \etal,\cite{ZhJeWh99} the HLF approximation
gives an overestimated shift of the equilibrium position of the oscillator in
the presence of an electron, and does not reproduce the retardation effects
when an electron hops onto a previously unoccupied site. Nevertheless, the
local lattice distortion at the site of the electron contains the crucial
impact of the electron on the lattice. Consequently, the transformed
Hamiltonian~(\ref{eq:LFHamiltonian}) should be a good starting point to
perform QMC simulations, which merely need to simulate small fluctuations
around the zero-point motion. Indeed we will see in Sec.~\ref{sec:resQMC}
that the expectation values of the phonon operators in the transformed
Holstein model remain close to the results of the free-oscillator case over
the whole range of the electron-phonon coupling. This makes sampling of the
phonon degrees of freedom much more efficient. In principle, it would also be
possible to develop a QMC algorithm starting with
Hamiltonian~(\ref{eq:htilde}), with the parameters $\gamma_{ij}$ being
determined by the variational method discussed in the following section. However,
we will see that the simple (local) Lang-Firsov transformation is already
sufficient to obtain a very efficient QMC method.

From Eq.~(\ref{eq:LFHamiltonian}) it is obvious that the standard Lang-Firsov
transformation on the one hand removes the electron-phonon coupling term, but
on the other hand introduces complex valued hopping integrals which depend on
the phonon momenta at the lattice sites involved in the hopping process.
Moreover, for more than one electron in the system, the last term $Q$
introduces a Hubbard-like attractive interaction.  In the case of the
extended transformation, the electron-phonon interaction term cannot be
eliminated entirely, the hopping term involves all phonon momenta $p_i$ as
well as the parameters $\gamma_{ij}$, and the electron-electron interaction
becomes long ranged.  For these reasons it is expedient to base the QMC
simulation on the local Lang-Firsov transformation~(\ref{eq:LFHamiltonian}).

For simplicity, we restrict ourselves in the present study to the case of a
single electron so that $\on_i\on_j=\on_i\delta_{ij}$. The electron-electron
interaction term in Hamiltonian~(\ref{eq:htilde}) becomes
\begin{equation}\label{eq:Iee1}
  \tilde{I}_\text{ee}=
  \sum_i\,\on_i\,\left(\frac{\om}{2}\sum_l\gamma^2_{li}-\alpha\gamma_{ii}\right)\,,
\end{equation}
while the corresponding term in the local Lang-Firsov transformation [last
term of Hamiltonian~(\ref{eq:LFHamiltonian})] reduces to
\begin{equation}\label{eq:Ep}
  Q\rightarrow -\lambda W/2\ = -E_\text{P}\,.
\end{equation}
Equations~(\ref{eq:Iee1}) and~(\ref{eq:Ep}) both describe a shift in energy
resulting from the original electron-electron interaction which is usually
called the polaron binding energy $E_\text{P}$.

\section{\label{sec:VPA}Variational Polaron Approach}

Here we present a simple variational method which is based on the extended
transformation discussed in the preceding section. Similar work along these lines
using different transformations of the Hamiltonian as well as physically
motivated wave functions can be found, for example, in
Refs.~\onlinecite{CaFiIa01},~\onlinecite{AcCuNoRo01}, and~\onlinecite{Ba02,Ba03}. As discussed above, the
zero-phonon ansatz of the simple HLF approximation gives reliable results
only in the limit of large $\omb$ and $\lambda$.  Whereas in HLF the
parameter $\gamma$ of the Lang-Firsov transformation is chosen such that the
electron-phonon coupling term $\tilde{I}_\text{ep}$ vanishes, in the
variational polaron approach (VPA), we treat the $\gamma_{ij}$ as variational
parameters which are determined by minimizing the ground-state energy in a
zero-phonon basis. Like the HLF approximation, the VPA becomes exact in the
weak-coupling limit $\lambda\rightarrow0$ and in the nonadiabatic
strong-coupling limit $\lambda,\omb\rightarrow\infty$. We will see in
Sec.~\ref{sec:resVPA} that the VPA also gives very accurate results for large
phonon frequencies, $\omb\gg1$.  This can easily be understood keeping in
mind the discussion of the validity of the HLF approximation given in the
preceding section. While the HLF ansatz overestimates the displacement of the
local oscillator in the presence of an electron, the VPA determines this
shift variationally. Moreover, the missing retardation effects in the
response of the oscillator to an electron hopping onto the site become
negligible for large phonon frequencies. Therefore, in addition to the cases
stated above, the VPA also becomes exact in the nonadiabatic limit
$\omb\rightarrow\infty$. Although the limitations of the VPA in or near the
adiabatic regime will clearly emerge when we discuss results in
Sec.~\ref{sec:resVPA}, it works surprisingly well if we keep in mind the
simplicity of the method. Moreover, the reasons for the failure of the VPA in
certain parameter regimes are physically clear and can easily be interpreted.

For translationally invariant systems the {\em displacement fields} satisfy
the condition $\gamma_{ij}=\gamma_{|i-j|}$.  Inserting this relation into
Eq.~(\ref{eq:Iee1}), the expression inside the brackets becomes independent
of the index $i$.  For the single electron case with $\sum_i \on_i=1$ we have
\begin{equation}
  \tilde{I}_\text{ee}=\frac{\om}{2}\sum_l\gamma^2_l-\alpha\gamma_0\,.
\end{equation}

We solve the eigenvalue problem of the transformed
Hamiltonian~(\ref{eq:htilde}) in a zero-phonon basis for which we make the ansatz
\begin{equation}\label{eq:basis}
  \ket{l} = c^\dag_{l\si}\ket{0} \otimes\prod_\nu \sket{\phi_0^{(\nu)}}
  \,,\quad l=1,\dots,N \,,
\end{equation}
where $\sket{\phi_0^{(\nu)}}$ denotes the ground state of the harmonic
oscillator at site $\nu$.  For simplicity, we restrict ourselves to one
dimension, although the method can easily be extended to higher dimensions.
The matrix elements of the hopping term in this basis are
\begin{eqnarray}\label{eq:K}\nonumber
   \bra{l}\tilde{K}\ket{l'} &=&
   -t_{ll'}\prod_\nu \sbra{\phi^{(\nu)}_0}
   e^{\rmi(\gamma_{l\nu}-\gamma_{l'\nu})\op_\nu}
   \sket{\phi^{(\nu)}_0}\\\nonumber
   &=& -t_{ll'}\;\prod_\nu\;\int\,\rmd x\, \phi(x+\gamma_{l\nu})\phi(x+\gamma_{l'\nu})\\
   &=&
   -t_{ll'} e^{-\frac{1}{4}\sum_\nu(\gamma_{\nu}-\gamma_{\nu+l-l'})^2}\,,
\end{eqnarray}
where $t_{ll'}= t\delta_{\las ll'\ras}$ is nonzero for nearest-neighbor
hopping pairs $l'=l\pm1$ and $\phi(x)$ is the harmonic-oscillator
ground-state wave function in coordinate space. The matrix elements of the
other terms of Hamiltonian~(\ref{eq:htilde}) are easily evaluated yielding
\begin{eqnarray}\label{eq:matelem}\nonumber
  \bra{l}P\ket{l'} &=& \delta_{ll'}\frac{\om}{2}\,,\\\nonumber
  \bra{l}\tilde{I}_\text{ep}\ket{l'} &=& 0\,,\\
  \bra{l}\tilde{I}_\text{ee}\ket{l'} &=&
  \delta_{ll'}\bigg(\frac{\om}{2}\sum_l\gamma_l^2
  -\alpha\gamma_0\bigg)\,.
\end{eqnarray}
The eigenstates of the transformed Hamiltonian~(\ref{eq:htilde}) in the
zero-phonon subspace, spanned by the basis states defined in
Eq.~(\ref{eq:basis}), are
\begin{equation}\label{eq:eigen}
  \sket{\psi_k} = c^\dag_{k\si}\ket{0}\otimes\prod_\nu\sket{\phi_0^{(\nu)}}
\end{equation}
with energy
\begin{eqnarray}\label{eq:E(k)}\nonumber
  E(k) &=& E_\text{k}
  + \frac{\om}{2}N + \frac{\om}{2}\sum_l\gamma_l^2-\alpha\gamma_0\\
  E_\text{k}&=&-t\sum_{\delta=\pm1} e^{\rmi k\delta}
  e^{-(1/4)\sum_\nu(\gamma_\nu-\gamma_{\nu+\delta})^2}
  \,,
\end{eqnarray}
where $E_\text{k}$ denotes the kinetic energy of the electron. Defining the
Fourier-transformed parameters
$\tilde{\gamma}_q$ as
\begin{equation}\label{eq:FTgamma}
  \tilde{\gamma}_q = \frac{1}{\sqrt{N}}\sum_l e^{\rmi ql}\gamma_l
\end{equation}
and using ($\gamma_l \in \Rs$)
\begin{equation}
  \sum_\nu \gamma_\nu \gamma_{\nu+\delta}=
  \sum_q\tilde{\gamma}_q\tilde{\gamma}_{-q} e^{\rmi q\delta}
  =\sum_q\tilde{\gamma}_q^2\cos q\delta \;,
\end{equation}
the kinetic energy can be written as
\begin{eqnarray}\nonumber
  E_\text{k} &=& -t\sum_\delta
  e^{\rmi k\delta} e^{-(1/2)
  \sum_q(1-\cos q\delta)\tilde{\gamma}_q^2}\\\nonumber
  &=& \tilde{\epsilon}_0(k) e^{-(1/2)
    \sum_q(1-\cos q)\tilde{\gamma}_q^2}\\
  &=& \tilde{\epsilon}_\text{eff}(k)\,,
\end{eqnarray}
where $\tilde{\epsilon}_0(k)=-2t\cos k$ is the tight-binding dispersion in
one dimension. Using these results the ground-state energy finally becomes
\begin{equation}\label{eq:evs}
  E(k) =\tilde{\epsilon}_\text{eff}(k) + \frac{N\om}{2}
  +\frac{\om}{2}\sum_q\tilde{\gamma}_q^2 -
  \frac{\alpha}{\sqrt{N}}\sum_q\tilde{\gamma}_q\,.
\end{equation}
The variational parameters $\tilde{\gamma}_p$ are determined by
\begin{equation}\label{eq:deriv}
\frac{\partial E}{\partial \tilde{\gamma}_p} =
-\tilde{\gamma}_p \tilde{\epsilon}_\text{eff}(p)
  (1-\cos p) + \om \tilde{\gamma}_p -
  \frac{\alpha}{\sqrt{N}} \overset{!}{=}0\,.
\end{equation}
The values for $\tilde{\gamma}_p$ which minimize the energy $E$ can then be
obtained from
\begin{equation}\label{eq:gammap}
  \tilde{\gamma}_p = \frac{g}{\sqrt{N}}
  \frac{1}{\om +
  \tilde{\epsilon}_\text{eff}(p)(1-\cos p)}\,.
\end{equation}
As $\tilde{\epsilon}_\text{eff}$ depends on the set of parameters
$\tilde{\gamma}_p$, this equation has to be solved self-consistently.
Equation~(\ref{eq:gammap}) has a typical random-phase approximation form, which is reasonable since a
variational ansatz for the wave function of the untransformed Hamiltonian can
be written as [see also Eq.~(\ref{eq:trop})]
\begin{equation}\label{eq:rpa_type}
 U^\dag\sket{\psi_k}
 = \frac{1}{\sqrt{N}} \sum_j
 e^{\rmi k j} \,c^\dag_{j\si} \,
 e^{-\rmi \sum_l \gamma_{jl} \hat p_l}\,
 \sket{0}\otimes\prod_\nu \sket{\phi_0^{(\nu)}}\,.
\end{equation}
In addition to the total energy given by Eq.~(\ref{eq:evs}), we are also
interested in the quasi-particle weight for momentum $k=0$, defined as
\begin{equation}\label{eq:def_z0}
  \sqrt{z_0} = \bra{0} \tilde{c}_{k=0,\si}\ket{\psi_0}\,.
\end{equation}
Here $\ket{\psi_0}$ denotes the ground state with one electron of momentum
$p=0$ and the oscillators in the ground state $\ket{\phi_0}$. Fourier
transformation leads to
\begin{eqnarray}\label{eq:z0}\nonumber
  \sqrt{z_0} &=& \frac{1}{N}\sum_{ij}\bra{\phi_0}\bra{0}
  \tilde{c}_{i\si}c^\dag_{j\si}\ket{0}\ket{\phi_0}\\\nonumber
  &=& \frac{1}{N}\sum_i \bra{\phi_0} e^{-\rmi\sum_k\gamma_{ik}
  \op_k}\ket{\phi_0}\\
  &=&  e^{-(1/4)\sum_q\tilde{\gamma}_q^2}\,,
\end{eqnarray}
where we have used the same steps as in Eq.~(\ref{eq:K}).

Results obtained with the VPA will be presented in
Sec.~\ref{sec:resVPA}.

\section{\label{sec:LFQMC}Monte Carlo for the transformed model}

In contrast to the approximate variational approach presented in the preceding
section, the QMC method discussed here is based on the exact Lang-Firsov
transformation of the Holstein Hamiltonian. Therefore, the method is exact
apart from statistical errors and Trotter discretization, as discussed in
Sec.~\ref{sec:introduction}.

\subsection{Partition function}\label{sec:QMCI}

We begin with the evaluation of the partition function $\Z=\tr e^{-\beta
  H}=\tr e^{-\beta\tilde{H}}$, with $\tilde{H}$ given by
Eq.~(\ref{eq:LFHamiltonian}). As indicated in the preceding section, for the case
of a single electron, the last term in Hamiltonian~(\ref{eq:LFHamiltonian})
represents a constant energy shift.  Moreover we can drop spin indices and
are left with the Hamiltonian
\begin{equation}
  \tilde{H} = \tilde{K} + P - E_\text{P}\,.
\end{equation}
The polaron binding energy given by Eq.~(\ref{eq:Ep}) can be neglected during
the QMC simulation, and needs only to be considered in calculating the total
energy. For simplicity, we only consider the one-dimensional case here,
although the generalization to higher dimensions is a simple matter.  Using
the Suzuki-Trotter decomposition we obtain\cite{wvl1992}
\begin{equation}\label{eq:suzuki-trotter}
  e^{-\beta\tilde{H}} \approx
  (e^{-\dtau\tilde{K}} e^{-\dtau P_\text{p}}
  e^{-\dtau P_\text{x}})^L \equiv \UU^L\,,
\end{equation}
where $\beta=(k_\text{B}T)^{-1}$ and $\dtau=\beta/L$. Splitting up the trace
into a bosonic and a fermionic part and inserting $L$ complete sets of
momentum eigenstates\cite{wvl1992} we derive the approximation for the
partition function
\begin{equation}
\Z_L =   \tr_\text{f}\int\,\rmd p_1\rmd p_2\cdots\rmd p_L
  \bra{p_1}\UU\ket{p_2}\cdots\bra{p_L}\UU\ket{p_1}\;,
\end{equation}
where $\rmd p_\tau\equiv\prod_i \rmd p_{i,\tau}$.  Each matrix element can be
evaluated by inserting a complete set of phonon coordinate eigenstates
$\int\rmd x_\tau \sket{x_\tau}\sbra{x_\tau}$. All $x_\tau$ integrals are of
Gaussian form and can easily be carried out. The result is
\begin{eqnarray}\nonumber
  \bra{p_\tau} e^{-\dtau P_\text{x}}\ket{p_{\tau+1}}
  &=&
  C^N
  e^{-\sum_i\left(p_{i,\tau}-p_{i,\tau+1}\right)^2/(2\om\dtau)}\,,\\
  C &=& \sqrt{\frac{2\pi}{\om\dtau}}\,.
\end{eqnarray}
The normalization factor in front of the exponential has to be taken into
account in the calculation of the total energy but cancels when we measure
other observables. With the abbreviation $\mathcal{D}p=\rmd p_1\rmd
p_2\cdots\rmd p_L$ the partition function finally becomes
\begin{equation}\label{eq:Z}
  \Z_L
   = C^{NL}
  \int\,\mathcal{D}p\,\, \wb \,\wf
\end{equation}
with the abbreviations
\begin{equation}\label{eq:omega}
   \wb=e^{-\dtau S_\text{b}} , ~~
   \wf=\text{Tr}_\text{f}\Omega ,~~
   \Omega=\prod_\tau e^{-\dtau\tilde{K}_\tau} \,.
\end{equation}
Here $\tilde{K}_\tau$ is $\tilde{K}$ with the phonon operators $\op_i$
replaced by the momentum $p_{i,\tau}$ on the $\tau$th Trotter slice.  The
exponential of the hopping term may in the single-electron case be written as
\begin{eqnarray}\label{eq:matrices}
  e^{-\dtau\tilde{K}_\tau} &=& D_\tau \kappa D_\tau^\dag\\\nonumber
  \kappa_{jj'} &=& \left(e^{\dtau t\,h^\text{tb}}\right)_{jj'}\,,\quad
  (D_\tau)_{jj'} = \delta_{jj'}e^{\rmi\gamma p_{j,\tau}}\,,
\end{eqnarray}
where $h^\text{tb}$ is the $N\times N$ tight-binding hopping matrix. Thus we
have the same matrix $\kappa$ for every time slice, which is transformed by
the diagonal unitary matrices $D_\tau$. In our one-electron case, the
fermionic weight $\wf=\sum_n \bra{n}\Omega\ket{n}$ is given by the sum over
the diagonal elements of the matrix representation of $\Omega$ in the basis
of one-electron states
\begin{equation}\label{eq:oneelbasis}
\ket{n}=c^\dag_n\ket{0}\,.
\end{equation}
The bosonic action in Eq.~(\ref{eq:omega}) contains only classical variables
and takes the form
\begin{equation}\label{eq:action}
  S_\text{b} =\frac{\om}{2}\sum_{i,\tau}p_{i,\tau}^2 +
  \frac{1}{2\om(\dtau)^2}\sum_{i,\tau}
  \left(p_{i,\tau} - p_{i,\tau+1}\right)^2\,,
\end{equation}
where the indices $i=1,\dots,N$ and $\tau=1,\dots,L$ run over all lattice
sites and Trotter times, respectively, with the periodic boundary conditions
$p_{i,L+1}=p_{i,1}$. It may also be written as
\begin{equation}\label{eq:action-w-matrix}
  S_\text{b} = \sum_i \bm{p}_i^\text{T} A \bm{p}_i
\end{equation}
with $\bm{p}_i=(p_{i,1},\dots,p_{i,L})$ and a ``periodic'' tridiagonal $L\times
L$ matrix $A$ with nonzero elements
\begin{equation}\label{eq:matrixA}
  A_{l,l} =\frac{\om}{2}+\frac{1}{\om\dtau^2}\;,\quad
  A_{l,l\pm1}=-\frac{1}{\om\dtau^2}\,.
\end{equation}
Since $\Z_L$ is a trace, it follows that  $A_{1,L}=A_{L,1}=-1/(\om\dtau^2)$.

At this stage, with the above result for the partition function, a QMC
simulation of the transformed Holstein model would proceed as follows. In
each MC step, a pair of indices $(i_0,\tau_0)$ on the $N\times L$ lattice of
phonon momenta $p_{i,\tau}$ is chosen at random. At this site, a change
$p_{i_0,\tau_0}\mapsto p_{i_0,\tau_0}+\Delta p$ of the phonon configuration
is proposed. To decide upon the acceptance of the new configuration using the
Metropolis algorithm, the corresponding weights $\wb\wf$ and $\wb'\wf'$ have
to be calculated. Due to the local updating process, the change of the
bosonic weight $\Delta\wb=\wb'-\wb$ can easily be obtained. In contrast, the
fermionic weight requires the evaluation of the $L$ fold matrix product
appearing in the definition of $\Omega$ in Eq.~(\ref{eq:omega}). The
numerical effort for the calculation of $\wf$ may be reduced by varying
$\tau_0$ sequentially from 1 to $L$ instead of picking random values. In this
case the calculation of the new fermionic weight, after the change of a
single phonon momentum, can be reduced to only two matrix multiplications.
Similar to other MC methods, a warm-up phase at the beginning of the
simulation would be required for each set of parameters.  An additional
difficulty arises from the fact that, for the transformed model, the
fermionic weight $\wf$ is no longer strictly positive, even for the case of a
single electron in one dimension. This is a consequence of the complex-valued
hopping integrals, in contrast to simulations of, \eg, the Hubbard model,
where a minus-sign problem occurs as a consequence of the Fermi statistics of
the electrons.\cite{wvl1992} Here the average sign of $\wf$ is smallest in
the regime of small phonon frequency and low temperature. The sign problem is
most pronounced for intermediate values of the electron-phonon coupling
strength $\lambda$, where the cross over from a large to a small polaron
occurs. However, in one dimension, it is not severe and reduces with
increasing system size. For example, calculations in the most
critical regime $\beta t=10$, $\omb=0.1$, and $\lambda\approx1$ have shown that
$\las\text{sign}\ras\equiv\las\wf\ras/\las|\wf|\ras$ increases from about 0.5
for $N=4$ to about 0.85 for $N=16$. Nevertheless, it remains to be seen to
what extent the number of electrons and the dimensionality of the system
affect the situation.

A related QMC approach to the original Holstein
Hamiltonian~(\ref{eq:holstein}) involves a very similar
derivation\cite{BlScSu81,ScSu81} to obtain the partition function, also in
the one-electron limit. In fact the bosonic action $S_\text{b}$ takes exactly
the same form, with $p$ replaced by $x$. The main difference is the fermionic
part of the partition function, contained in the matrix $\Omega$. While the
Lang-Firsov transformation leads to a complicated hopping term, the standard
approach for the untransformed model only includes the bare hopping operator
given by Eq.~(\ref{eq:holstein}). However, an interaction term $I$ [cf.
Eq.~(\ref{eq:holstein})] appears, which contains the phonon coordinate $\ox$.
Hence the matrix $\Omega$ is replaced by
\begin{equation}
  {\Omega'} = \prod_{\tau=1}^L \kappa\,V_\tau\,,\quad
  (V_\tau)_{jj'}=\delta_{jj'}\,e^{\dtau\alpha x_{j,\tau}}
\end{equation}
and the path integral in the partition function [Eq.~(\ref{eq:Z})] is over
all coordinates $x$ instead of the momenta $p$.  Apart from the fact that the
coordinates $x$ are sampled instead of the phonon momenta, the QMC procedure
for the untransformed model is identical to the simulation described above.
We shall refer to this less sophisticated QMC method for the original
Holstein Hamiltonian as the {\it standard approach.}  For $\lambda=\alpha=0$,
\ie, no electron-phonon coupling, we have a set of $N$ independent harmonic
oscillators, and both approaches are alike.

\subsection{Problems with the standard approach}\label{sec:QMCproblems}

Let us briefly consider the noninteracting limit, in which the partition
function can be written as $\Z_L\sim\int\mathcal{D}p\,e^{-\dtau
  S_\text{b}}$.  As discussed by Batrouni and Scalettar,\cite{batscal} the
difficulties encountered in QMC simulations, even for the simple case of a
single ($N=1$) harmonic oscillator, arise from the large condition number,
\ie, the ratio of largest to smallest eigenvalue, of the bosonic action
$S_\text{b}$. For small values of $\dtau$ this ratio is proportional to
$(\om\dtau)^{-2}$ (Ref.~\onlinecite{batscal}), leading to autocorrelation
times which grow quadratically with decreasing phonon frequency and the
number $L$ of Trotter times.  The physical reason for these correlations
becomes obvious if we look at the bosonic action [Eq.~(\ref{eq:action})]. The
latter can be thought of as being proportional to the energy of a given
phonon configuration, $E=\dtau S_\text{b}$. While the first term corresponds
to the kinetic energy of the oscillators, the second term describes a
coupling in imaginary time, \ie, a pure quantum effect. As pointed out by
Batrouni and Scalettar,\cite{batscal} large changes of a single phonon degree
of freedom, $p_{i,\tau}$ say, are very unlikely to be accepted due to the
energy change proportional to $1/(\om\dtau)$, which arises from the coupling
to $p_{i,\tau\pm1}$. However, a QMC simulation with only small local changes
is extremely ineffective in sampling the relevant regions of phase space.
Therefore, successive phonon configurations will be highly correlated.  A
possible solution might be the use of global updating schemes. Alternatively,
the situation could be improved by transforming to the normal modes of the
phonons, so that different step sizes can be used in updating each mode. We
will see in Sec.~\ref{sec:princ-comp-repr} that the principal component
representation can indeed be used to completely eliminate these difficulties.

In addition to the above-mentioned autocorrelations, which are in fact
independent of any electronic influences, standard simulations of the
Holstein model become very difficult in the regime where polaron effects are
large. This occurs at low temperatures, small phonon frequencies $\omb<1$, and
for intermediate or strong electron-phonon coupling $\lambda\gtrsim1$.
Unfortunately, these are exactly the parameters of interest for simulations
of real substances such as, \eg, the manganites.\cite{David_AiP}  To
illustrate the physical origin of these problems let us consider the case of
a single electron in the Holstein model. As discussed in
Sec.~\ref{sec:holstein}, in the polaronic regime, the electron drags with it
a cloud of phonons which corresponds to a more or less localized lattice
distortion. When the electron hops from site A (with a displaced oscillator
corresponding to a small polaron), say, to a neighboring, previously
unoccupied site B (with the oscillator in its undisplaced ground state)
during a QMC simulation, the current phonon configuration is no longer
energetically favorable. Clearly, the oscillator at site A has to return to
its undisplaced ground state, while a corresponding phonon cloud has to be
built up at site B. Such distortions of the lattice in the presence of an
electron are large compared to the zero-point motion of the oscillator. On
the other hand, only small changes of the current configuration will be
accepted in the simulation. Consequently it takes an enormous number of
single updates to obtain the new configuration in which the polaron has
completely moved to site B. Obviously these polaron effects also give rise to
strongly autocorrelated configurations, thereby dramatically increasing the
numerical effort for the simulation. These problems due to polaron formation
can be overcome by using the Lang-Firsov transformed model. The
transformation separates the large displacements of the local oscillators,
due to polaron effects, from the free-oscillator dynamics which correspond to
vibrations around the shifted equilibrium positions. The quantities to be
sampled, namely the phonon momenta $p$, only show a weak dependence on the
{electron-phonon} coupling strength $\lambda$, in stark contrast to the
coordinates $x$ in the original, untransformed model, whose expectation
values grow linearly with $\lambda$ in the strong-coupling regime. In fact,
the QMC results obtained for the transformed model (see also
Sec.~\ref{sec:resQMC}) show that the statistical errors increase in the
intermediate coupling regime $\lambda\approx 1$, but decrease again as we
approach the strong-coupling limit. This is in perfect agreement with the
fact that the the Lang-Firsov transformation diagonalizes the
Hamiltonian~(\ref{eq:holstein}) in the strong-coupling or atomic limit
$\lambda\rightarrow\infty$ (see Sec.~\ref{sec:lang-firs-transf}), so that the
QMC method based on the transformed model becomes more and more efficient as
$\lambda$ increases.

\subsection{Observables}

Thermodynamic expectation values
\begin{equation}
  \las O\ras = \Z^{-1}\,\tr\, \hat{       O} \,e^{-\beta        H}
             = \Z^{-1}\,\tr\, \hat{\Tilde{O}}\,e^{-\beta \tilde{H}}
\end{equation}
of observables $O$ are computed in the Lang-Firsov transformed representation
via 
\begin{equation}
  \las O\ras = \Z^{-1}\,\tr_\text{f} \int\,\rmd p\,
  \bra{p}\hat{\Tilde{O}}\,e^{-\beta\tilde{H}}\ket{p}\,.
\end{equation}
In this paper we are interested in the kinetic energy of the electron, the
total energy, the mean square of the phonon momenta, and the momentum
distribution $n(k)\equiv\las\tilde{c}^\dag_k\tilde{c}_k\ras$ for various
wave vectors $k$. We begin with the kinetic energy which is defined as
\begin{equation}\label{eq:Ekin}
  E_\text{k} = \las K \ras
             = -t \Z^{-1}\,
             \sum_{\las ij\ras}\text{Tr}\;\big( \,c_i^\dag c_j\,
  e^{\rmi\gamma(\op_i-\op_j)} \,e^{-\beta \tilde{H}}\, \big)\,.
\end{equation}
Using the same steps as in the derivation of the partition function (see
Sec.~\ref{sec:QMCI}), and absorbing the additional phase factor in a matrix
$M=D_1^\dag\Omega D_1$ [see Eq.~(\ref{eq:matrices})], we find
\begin{eqnarray}\nonumber
  E_\text{k}
  &=&
   -t \Z^{-1}_L\,\sum_{\las ij\ras}\int\,\D p\,\wb\sum_n
  \bra{n}M c^\dag_i c_j\ket{n}\\\nonumber
  &=&
   -t \Z^{-1}_L\,\sum_{\las ij\ras}\int\,\D p\,\wb
  \bra{j}M\ket{i}
\end{eqnarray}
with one-electron states $\ket{n}$ as defined in Eq.~(\ref{eq:oneelbasis}).
Using the matrix elements $M_{ij} = \bra{i}M\ket{j}$ and the expectation
values
\begin{equation}\label{eq:Ob}
\las O \ras_\text{b} = \frac{\int\,\D p\,\wb\; O(p)}{ \int\,\D p\,\wb}
\end{equation}
with respect to the purely phononic weights $\wb$ we obtain
\begin{equation}\label{eq:Ek}
  E_\text{k}=
  -t\; \frac{\sum_{\las ij\ras}\;\las M_{ji}\ras_\text{b}}
  {
    \sum_{i}\;\las M_{ii}\ras_\text{b}
  }\,.
\end{equation}
Here we have already taken into account the reweighting method which will be
discussed in detail in the following section.  The total energy can be
obtained from the thermodynamic relation $E=-\partial(\ln\Z)/\partial\beta$,
with $\mathcal{Z}$ given by Eq.~(\ref{eq:Z}). The result is
\begin{eqnarray}\label{eg:E0}
  E &=& E_\text{k} + \frac{\om}{2}\sum_i\la p_i^2\ra
  + E'_\text{ph} - E_\text{P}\,,\\\nonumber
  E'_\text{ph}&=&\frac{N}{2\dtau}-\frac{1}{2\om\dtau^2L}
  \sum_{i,\tau}\la\left(p_{i,\tau}-p_{i,\tau+1}\right)^2\ra
  \,,
\end{eqnarray}
where $E_\text{P}$ is defined in Eq.~(\ref{eq:Ep}) and the expectation values
are calculated according to Eq.~(\ref{eq:reweighting}) given below. To
compare with other work we subtract the ground-state energy of the phonons,
$E_{0,\text{ph}}=N\om/2$. Finally, $n(k)$ can be obtained using Fourier
transformation. The result is
\begin{equation}\label{eq:nk}
  n(k)= \frac{1}{N}\frac{\sum_{ij}\la M_{ij}\ra_\text{b} e^{\rmi k(i-j)}}
        {\sum_{i} \las M_{ii}\ras_\text{b}}
\end{equation}
with $k$ from the first Brillouin zone and the same matrix $M$ as in the case
of the kinetic energy.

\section{\label{sec:reweighting}Reweighting}

In typical QMC simulations a large amount of the total computational effort
goes into the calculation of the probability for the acceptance of a proposed
change of the configuration.  This probability is usually determined by the
ratio of the weights of the new and the old configuration, as in the
Metropolis algorithm used here. In the notation of Sec.~\ref{sec:LFQMC}, this
involves the calculation of $\wb$ and $\wf$ for the two configurations, $S$
and $S'$ say, in every MC step.  While the change in the
bosonic weight, $\wb(S')/\wb(S)$, is easily calculated for the case of local
updating, the fermionic weight given by Eq.~(\ref{eq:omega}) involves an
$L$-fold matrix product of $N\times N$ matrices for each configuration.
Although the numerical effort of the evaluation of such a matrix product can
be reduced by scanning sequentially through the time slices (see
Sec.~\ref{sec:QMCI}) it still requires a lot of total computer time.

This can be avoided by reweighting of the probability distribution to be
sampled. In the case considered here, this corresponds to taking into account
only the change $\wb(S')/\wb(S)$ in the bosonic weight, and compensating for
this by dividing the resulting expectation value by the expectation value of
the fermionic weight $\wf$, as has been used already in Eq.~\ref{eq:Ek},
leading generally to ratios of the form
\begin{equation}\label{eq:reweighting}
  \las O \ras ~=~
  \frac{\las O\, \wf\ras_\text{b}}{\las \wf\ras_\text{b}}\,,
\end{equation}
where the subscript ``b,'' defined in Eq.~(\ref{eq:Ob}), indicates that the
average is computed based on $\wb$ only. Following this procedure, the
fermionic weight is treated as part of the observables. The splitting into
weight $\wb$ and observable $O \wf$ is sensible as long as the variance of
$\wf$ and $O \wf$ is small, which is the case after the Lang-Firsov
transformation. This approach has several additional advantages. With the
reweighting method, the updating of the system does no longer require the
calculation of $\wf$ in every step, but only when measurements are performed.
Compared to the usual QMC procedure described in Sec.~\ref{sec:QMCI}, this
can save an enormous amount of computer time, allowing such simulations to be
run on a standard personal computer instead of a high-performance
supercomputer.  Additionally, since the updating does no longer involve any
electronic contributions, it becomes independent of the electron-phonon
coupling strength $\lambda$.  This allows the simultaneous measurement of
observables for a whole set of values of $\lambda$ in a single MC run.  For a
given phonon configuration, the fermionic weight and the observables are
measured and stored for each value of the coupling. This procedure is
repeated until the required number of measurements has been made. At the end
of the simulation an appropriate analysis of the measured values is made
independently for each $\lambda$.  In contrast, the QMC procedure without
reweighting (see Sec.~\ref{sec:QMCI}) would require a separate run for each
value of $\lambda$, including a warm-up phase to equilibrate the system for
the current set of parameters.  We will see in Sec.~\ref{sec:princ-comp-repr}
that in combination with the principal component representation, the phonon
momenta $p$ can be sampled exactly, removing all autocorrelations. This
avoids a warm-up phase, and measurements can be made after every Monte Carlo
step. In this final, very efficient procedure, the calculation of $\wf$ for
measurements remains, and is then the most time-consuming part of the
calculation.  Finally, we want to point out that, with the use of the
reweighting method, the electronic degrees of freedom are treated exactly,
\ie, they are not sampled in the course of the simulation.

As mentioned in Sec.~\ref{sec:QMCI}, the weight $\wf$ for the transformed
model is no longer strictly positive, so that it cannot be interpreted as a
probability. The usual way to deal with such a sign problem is to split the
weight into $\wf\equiv|\wf|\,\text{sgn}\,\wf$. Then $|\wf|$ can be used as
the weight of a given configuration in the updating process, while the sign
is absorbed in the observables. The difference to the reweighting method
presented here is that instead of the sign of $\wf$, we treat the whole
weight $\wf$ as part of the observables.

Despite the obvious advantages of this approach, it is necessary to
scrutinize whether reweighting does not lead to prohibitive statistical
noise.  If, for example, there was too small an overlap of the actual
probability distribution with the one we are sampling with, the method would
fail. In fact, our calculations have shown that for the {\em un}transformed
model the reweighting method would fail at low temperatures and for critical
values of the parameters $\omb$ and $\lambda$.

\begin{figure}
  \includegraphics[width=0.45\textwidth]{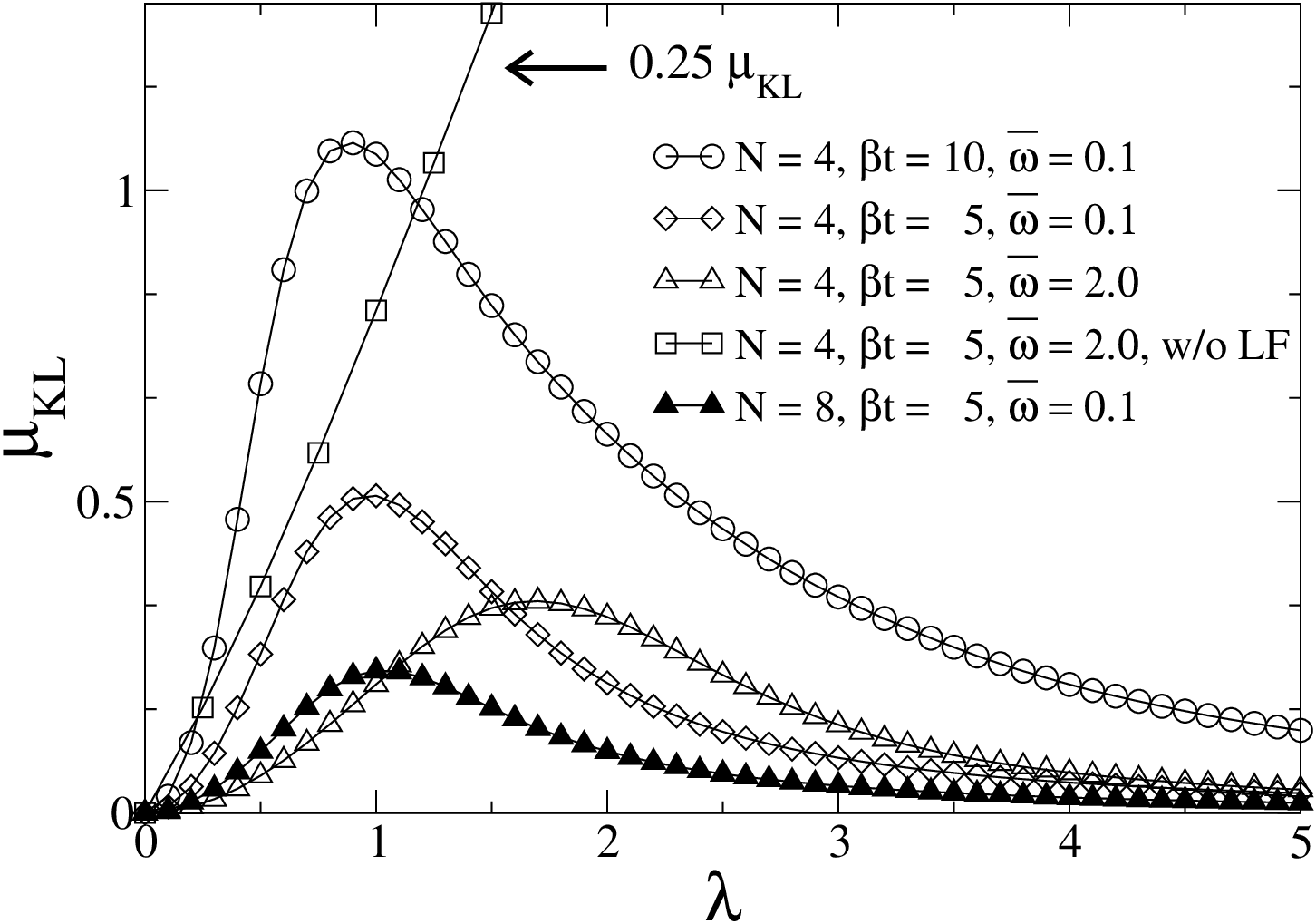}
\caption{\label{fig:mu_KL}
  Kullback-Leibler number $\mu_\text{KL}$ as a function of
  electron-phonon coupling $\lambda$ for various sets of the parameters $N$
  (number of sites), $\beta$ (inverse temperature) and $\omb$.  As indicated,
  the results for the untransformed model, denoted in the legend as ``w/o LF,''
  have been scaled by a factor $0.25$ (see text). Error bars are smaller than
  the symbols shown, and lines are guides to the eye only.}
\end{figure}

The distance between two arbitrary probability distributions $\phi_1(x)$ and
$\phi_2(x)$, each depending on a set of variables $x$, can be measured by the
so-called Kullback-Leibler number $\mu_\text{KL}$ which is defined
as\cite{KapKes92}
\begin{equation}\label{eq:KL}
  \mu_\text{KL}(\phi_1,\phi_2) = \int\,\rmd x\,\phi_1(x)\ln\frac{\phi_1(x)}
  {\phi_2(x)}\,.
\end{equation}
For $\phi_1\equiv\phi_2$ we have $\mu_\text{KL}=0$, while for
$\phi_1\neq\phi_2$ $\mu_\text{KL}>0$. The fact that $\mu_\text{KL}$ is a
reasonable measure for the distance of two distributions is best illustrated
by considering two Gaussian deviates $\phi_1$, $\phi_2$ with variance
$\si^2$, centered at $x_1$ and $x_2$, respectively. In this case
$\mu_\text{KL}=(x_1-x_2)^2/(2\si^2)$.  For $|x_1-x_2|=\sqrt{2}\si$, where the
two peaks begin to be distinguishable, we have $\mu_\text{KL}=1$, while a
large value of $\mu_\text{KL}\simeq10$, for example, corresponds to
well-separated Gaussian distributions.  Here we use the Kullback-Leibler
number to investigate the applicability of the reweighting method.  As long
as the Kullback-Leibler number is less than or comparable to 1, reweighting
works well, while a Kullback-Leibler number strongly exceeding unity
indicates severe problems.  Two relevant distributions in our case are given
by $\phi_1(\bm{p})=\wb(\bm{p})/\Z_\text{b}$ and
$\phi_2(\bm{p})=\wb(\bm{p})|\wf(\bm{p})|/\Z_\text{bf}$, depending on the
phonon configuration $\bm{p}$ (or $\bm{x}$ in the case of the untransformed
Holstein model). $\Z_\text{b}$ and $\Z_\text{bf}$ are the normalization
factors of the probability densities $\phi_1(\bm{p})$ and $\phi_2(\bm{p})$,
and $\wf$ has been replaced by its absolute value due to the aforementioned
sign problem. Inserting these definitions into Eq.~(\ref{eq:KL}) we find
$\mu_\text{KL}=\ln\las |\wf|\ras_\text{b}-\las\ln|\wf|\ras_\text{b}$.
Figure~\ref{fig:mu_KL} shows results for $\mu_\text{KL}$ for different
parameters $\beta$, $\omb$, and $N$.  For $\lambda=0$, $\wf$ is independent of
the phonon configuration so that $\mu_\text{KL}=0$. With increasing
electron-phonon coupling, the difference between the two distributions
becomes larger. For an intermediate value of the electron-phonon coupling
strength, $\mu_\text{KT}$ takes on a maximum and approaches zero again in the
strong-coupling limit $\lambda\rightarrow\infty$. This is exactly the
behavior we would expect for the Lang-Firsov transformed model. For
$\lambda=0$ the transformation has no effect and $\wf$ is a constant, just as
in the case of the untransformed model. In the intermediate coupling regime,
the small-polaron picture mediated by the transformation is not correct as we
have an extended (large) polaron in this region. However, as the coupling
increases further, the polaron becomes smaller and for $\lambda=\infty$ it is
known that the Lang-Firsov transformation diagonalizes the Holstein
Hamiltonian~(\ref{eq:holstein}). The dependence of $\mu_\text{KL}$ on the
temperature and the phonon frequency is also in perfect agreement with the
physical picture of the Holstein polaron. As $\beta t$ increases, polaron
effects become more prominent. The same effect occurs if we decrease $\omb$,
and in both cases the maximum of $\mu_\text{KL}$ increases. In
Fig.~\ref{fig:mu_KL} the result for a system of eight lattice sites is also
shown. The maximum in $\mu_\text{KL}$ is clearly smaller than for the
four-site cluster. Calculations for even larger clusters (not shown) reveal that the
maximum in $\mu_\text{KL}$ decreases further indicating that the overlap
between $\phi_1$ and $\phi_2$ increases as $N\rightarrow\infty$.
This behavior agrees well with the influence of finite-size effects in the
transition region as pointed out before by Marsiglio.\cite{Marsiglio95} As the
system size increases, the cross over becomes smoother in agreement with the
fact that the ground state of the Holstein polaron is an analytic function of
the coupling $\lambda$ (Ref.~\onlinecite{Loe88}). This point will be further
illustrated in Sec.~\ref{sec:resQMC}.  To summarize, for all parameters shown
in Fig.~\ref{fig:mu_KL}, the maximum of $\mu_\text{KL}$ lies below
$\mu_\text{KL}\approx1$, so that we can conclude that the two distributions
are indeed very close and the reweighting method can be successfully applied.

We have also calculated the Kullback-Leibler number for the case of the
untransformed model, denoted in Fig.~\ref{fig:mu_KL} as ``w/o LF,'' for which
$|\wf|\equiv\wf$. The result has been divided
by a factor 4 to allow a better representation in Fig.~\ref{fig:mu_KL}.
The difference between $\phi_1$ and $\phi_2$ increases strongly with
$\lambda$ and reaches large values of $\mu_\text{KT}>10$ already in the
intermediate coupling regime $1<\lambda<2$. Hence we cannot expect the
reweighting method to work in this case. Finally we want to point out that
the distance between $\phi_1$ and $\phi_2$ may not affect all observables in
the same way. A detailed analysis for each observable $O$ would be based on
the Kullback-Leibler distance of the marginal probability densities
\[
p_\alpha(o) = \int \rmd x\; p(o|x)\; p_\alpha(x) =
\int \rmd x\; \delta(o - O(x))\; p_\alpha(x)\;,
\]
where $O(x)$ is the value of the observable for a given configuration $x$ and
$\alpha=1,2$ for the two distributions under consideration.

In summary, the reweighting method, together with the Lang-Firsov
transformation, allows us to sample a system of independent oscillators,
while all the influence of the electrons is transferred to the observable,
thereby strongly reducing the numerical effort.  In order to obtain a
reliable error analysis for observables calculated according to
Eq.~(\ref{eq:reweighting}), the jackknife procedure\cite{DavHin} has been
applied.

\section{\label{sec:princ-comp-repr}Principal component representation}

Although the reweighting method allows us, in principle, to skip enough
sweeps between measurements to reduce autocorrelations to a minimum, the
computational effort for these Monte Carlo updates can become the most
time-consuming part of the simulation. Even though a single phonon update
requires negligible computer time compared to the evaluation of the fermionic
weight, in the critical parameter regime, an enormous number of such steps
will be necessary between successive measurements. Moreover, reliable results
can only be obtained when long enough Monte Carlo runs have been performed to
see even the longest autocorrelation times. In this section, we present a
principal component representation for the phonon degrees of freedom,
which enables us to create completely uncorrelated samples of phonon
configurations.

In order to illustrate the severe problem of autocorrelations with standard
updates of the phonons, we have calculated the integrated autocorrelation
time $\tau_\text{int}^p$ for the phonon momenta $p$. $\tau_\text{int}$ is a
direct measure for the number of MC steps which have to be skipped between
measurements in order to obtain uncorrelated data, and is usually given in
units of sweeps.  We define a sweep as $N$ times $L$ proposed local changes
of the phonon configuration.  For a four-site system, for example, with
$\beta t=5$, $\lambda=2$, $\omb=2$, and $\dtau=0.05$ we find
$\tau_\text{int}^p\approx 500$. This corresponds to an autocorrelation time
of about $2\times10^5$ single MC steps. For smaller phonon frequencies,
$\tau_\text{int}$ increases strongly. For $\omb=1$ and the same $\dtau$, the
autocorrelation time is already $\approx 1700$ sweeps, which agrees quite
well with the $(\om\dtau)^{-2}$ dependence of the correlations for
$\lambda=0$ given in Sec.~\ref{sec:QMCproblems}.  The dependence of
$\tau_\text{int}^p$ on the coupling strength $\lambda$ is relatively weak,
and we have found no systematic behavior of $\tau_\text{int}^p$ as a function
of $\lambda$.  Depending on the other parameters, the autocorrelation times
were observed to increase or even decrease slightly as $\lambda$ is
increased.  This behavior can be ascribed to the exact treatment of the
fermion degrees of freedom. As we are not sampling the hopping process of the
single electron considered here, no autocorrelations due to the resulting
reaction of the harmonic oscillators to the electronic motion (see
Sec.~\ref{sec:QMCproblems}) can occur. Moreover, even if we would sample the
electronic degrees of freedom in the QMC simulation, these autocorrelations
would still be strongly reduced as long as we use the Lang-Firsov transformed
model. This is a consequence of the fact that the large displacements of the
oscillators in the presence of an electron are explicitly contained in the
Hamiltonian~(\ref{eq:LFHamiltonian}).  Finally, as the number of lattice
sites is varied, $\tau_\text{int}^p$ remains constant in units of sweeps for
our single-electron simulations. We also determined the autocorrelation times
for observables such as, \eg, the kinetic energy. Although $\tau_\text{int}$
is smaller for electronic observables, the problem still exists, and the
determination of the autocorrelation times for various parameter sets is
vital to obtain reliable results. This usually requires very long QMC runs
and a lot of CPU time.

As indicated in Sec.~\ref{sec:QMCproblems}, the autocorrelations which arise
from the structure of the bosonic action $S_\text{b}$ [see
Eq.~(\ref{eq:action})] may be overcome by a transformation to the normal
modes of the system. Here we represent the bosonic action $S_\text{b}$ in
terms of its normal modes along the imaginary time axis. This allows us to
sample completely uncorrelated phonon configurations. In combination with the
reweighting method the fermion degrees of freedom are treated exactly, so
that our QMC method is indeed free of any autocorrelations.  This greatly
simplifies calculations, since it makes the usual binning analysis (to
determine the autocorrelation times) obsolete and, more importantly, leads to
significantly shorter simulation times.

All this can be achieved with the simple but effective idea of a
transformation to principal components (PCs). To this end let us recall the
form of the bosonic action given by Eq.~(\ref{eq:action-w-matrix}) which can
also be written as
\begin{equation}\label{eq:pc}
  S_\text{b}=
  \sum_i \bm{p}_i^\text{T} A \bm{p}_i =
  \sum_i \bm{p}_i^\text{T} A^{1/2}
         A^{1/2} \bm{p}_i=:
  \sum_i \bm{\xi}_i^\text{T}\cdot\bm{\xi}_i
\end{equation}
with the PCs $\bm{\xi}_i=A^{1/2}\bm{p}_i$, in terms of which the
bosonic weight takes the simple Gaussian form
\begin{equation}\label{eq:action_quad}
  \wb=e^{-\dtau\sum_i \bm{\xi}^\text{T}_i\cdot\bm{\xi}_i}\,.
\end{equation}
The QMC can now be performed directly in terms of the new variables $\xi$.
To calculate observables we have to transform back to the phonon momenta $p$
using the matrix $A^{-1/2}$. Comparison with Eq.~(\ref{eq:action-w-matrix})
shows that instead of the ill-conditioned matrix $A$ we now have the ideal
structure that we can easily generate exact samples of a Gaussian
distribution. In terms of the new coordinates $\xi$, the probability
distribution can be sampled exactly, e.g., by the Box-M\"uller
method.\cite{numrec_web} In contrast to a standard Markov chain MC
simulation, every new configuration is accepted, and measurements of
observables can be made at each step.

From the definition of the PCs it is obvious that an update of a single
variable $\xi_{i,\tau}$, say, actually corresponds to a change of all phonon
coordinates $p_{i,\tau'}$\,, $\tau'=1,\dots,L$.  Thus, in terms of the
original phonon coordinates $p_i$, the updating loses its local character.
As a consequence, the sequential updating of the Trotter time slices, which
we mentioned in Sec.~\ref{sec:LFQMC}, can no longer be exploited to reduce
the numerical effort for the evaluation of the fermionic weight. However, in
combination with the reweighting method, the latter is only calculated when
measurements of observables are made. The enormous advantage of the PCs,
leading to completely uncorrelated phonon configurations, clearly outweighs
this drawback. Nevertheless, this restriction has to be kept in mind when
considering possible extensions to many-electron systems. Apart from this,
the PC representation can also be applied to the more general case of more
than one electron, since the bosonic action [Eq.~(\ref{eq:action_quad}), on
which the transformation relies, remains unchanged]. This even holds for the
case of more general models including, \eg, spin-spin or Hubbard-type
interactions, as long as the phonon operators enter in the same form as in
the Holstein model.

Another important point is the combination of the PCs with the reweighting
method. Using the latter, the changes to the original momenta $p$, which are
made in the simulation, do not depend in any way on the electronic degrees of
freedom. Thus we are actually sampling a set of $N$ independent harmonic
oscillators, as described by the purely bosonic action $S_\text{b}$. The
crucial requirement for the success of this method is the use of the
Lang-Firsov transformed model, in which the polaron effects are separated
from the zero-point motion of the oscillators around their current
equilibrium positions.

Finally, for the {\em untransformed} model, Eq.~(\ref{eq:holstein}), the
bosonic action can be obtained from Eq.~(\ref{eq:action-w-matrix}) by
replacing $p$ with $x$ (see Sec.~\ref{sec:QMCI}) and a transformation to PCs
could also be used. However, as discussed in Sec.~\ref{sec:reweighting},
without the Lang-Firsov transformation, the reweighting procedure fails.
Consequently, using the standard approach, the phonon coordinates $x$ would
depend on the electronic degrees of freedom, and this makes exact sampling
impossible for the untransformed model.

\section{\label{sec:resVPA}Results: VPA}

\begin{figure}
  \includegraphics[width=0.45\textwidth]{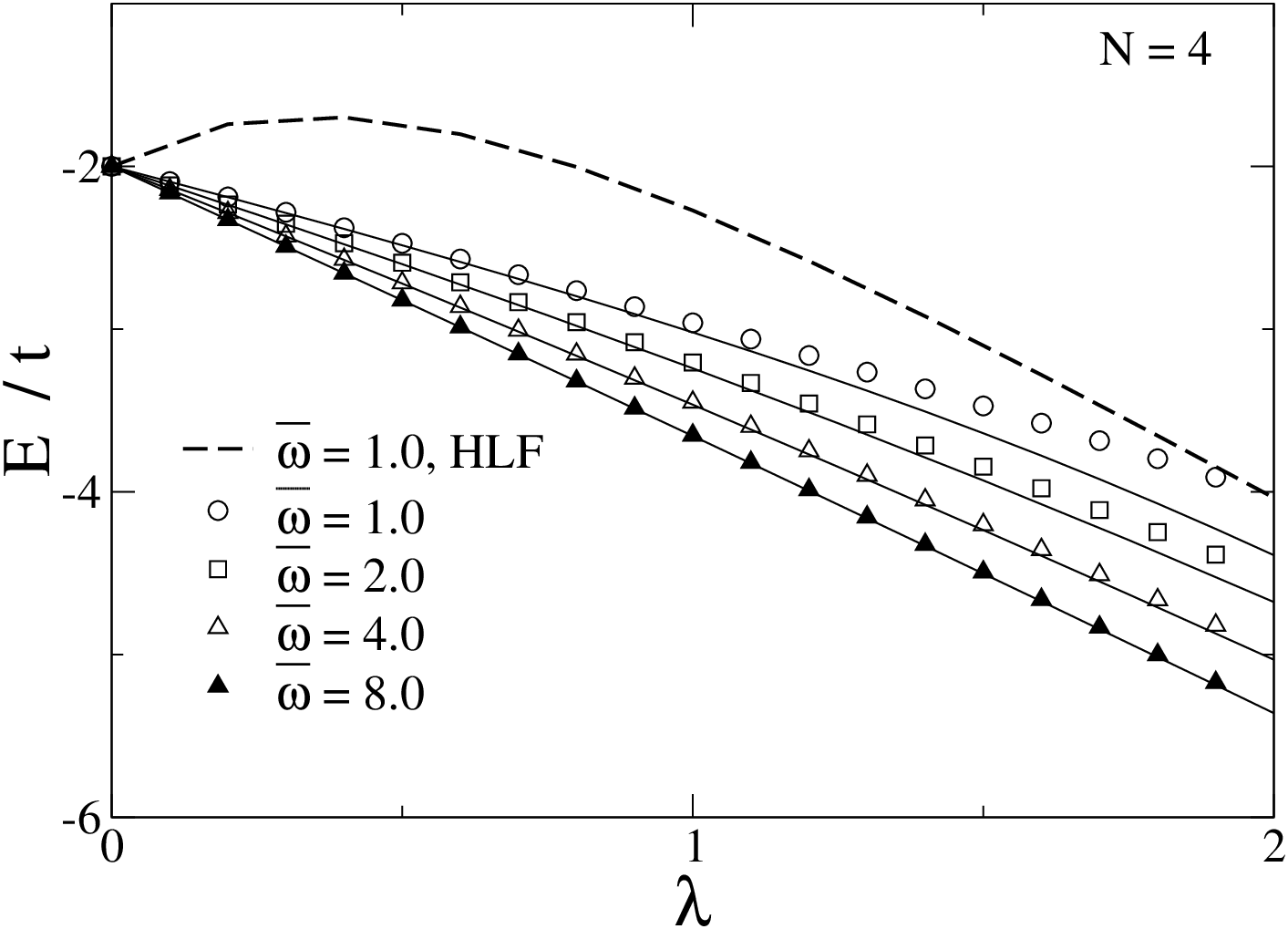}
  \includegraphics[width=0.45\textwidth]{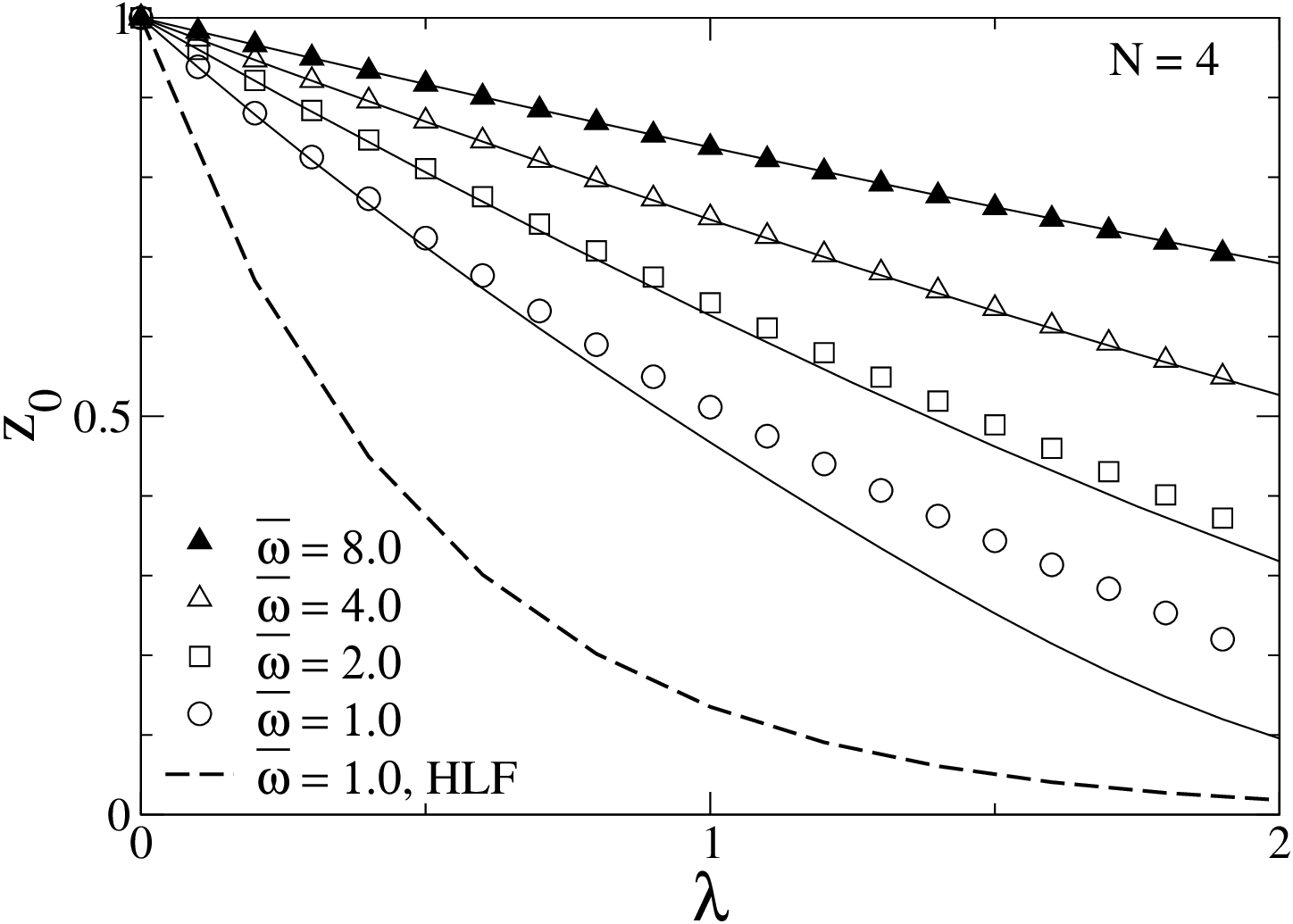}
\caption{\label{fig:E0z0vpa}
  Total energy $E$ (top) and quasiparticle weight $z_0$ (bottom) as 
  functions of the electron-phonon coupling $\lambda$ for different values of
  the phonon frequency $\omb$.  Symbols correspond to VPA results, while full
  lines represent exact $T=0$ data obtained with the Lanczos
  method (Ref.~\onlinecite{Mars95}). Dashed lines are results of the HLF approximation.}
\end{figure}
\begin{figure}
  \includegraphics[width=0.45\textwidth]{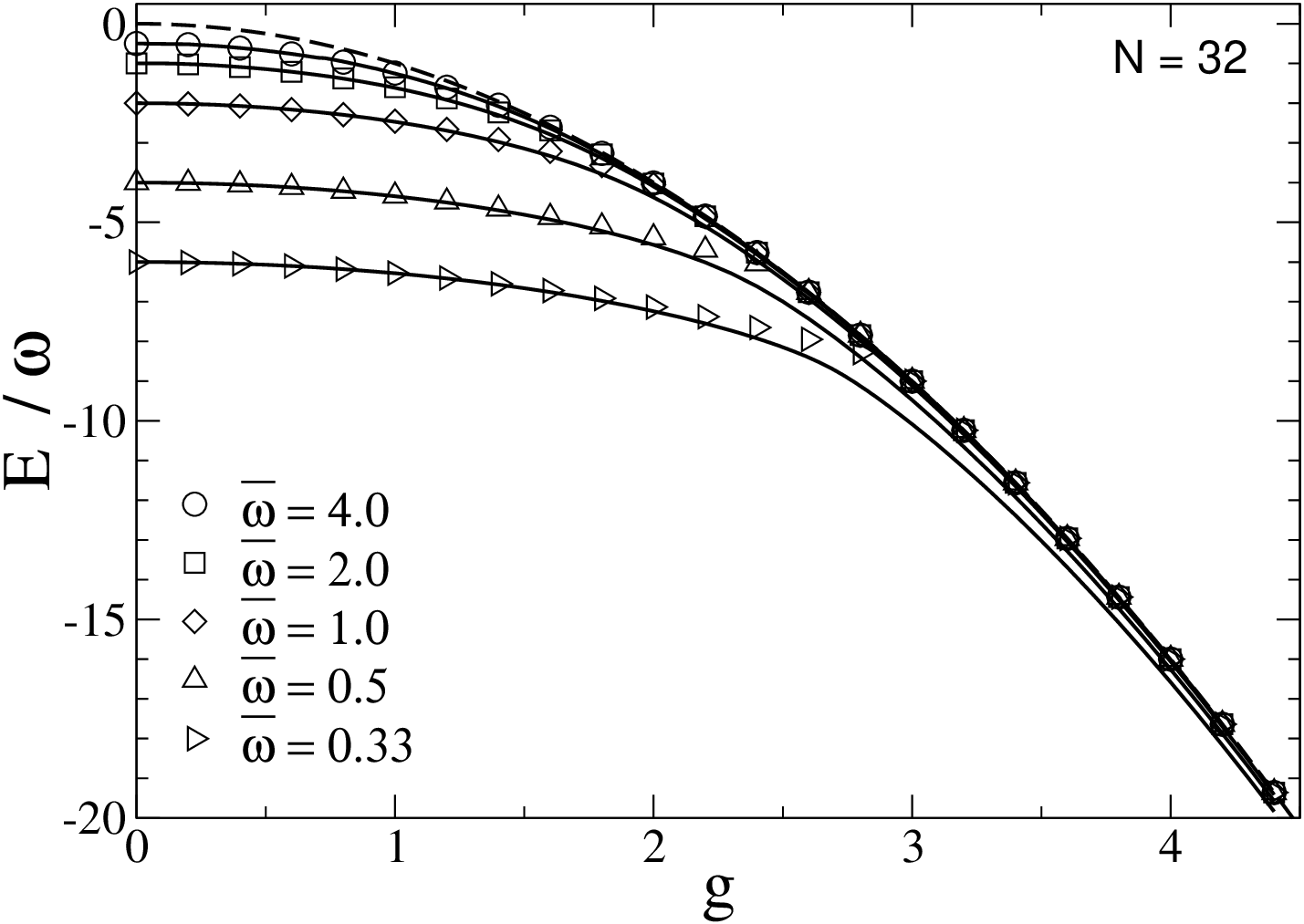}
\caption{\label{fig:E0vpa_GL}
  Total energy $E$ as a function of the electron-phonon coupling $g$ (see
  text) for different values of the phonon frequency $\omb$.  Symbols
  correspond to VPA results, while full lines represent data
  obtained with the globallLocal method (Ref.~\onlinecite{RoBrLi99II}). The
  dashed line represents the atomic-limit result ($\omb=\infty$).}
\end{figure}
\begin{figure}
  \includegraphics[width=0.45\textwidth]{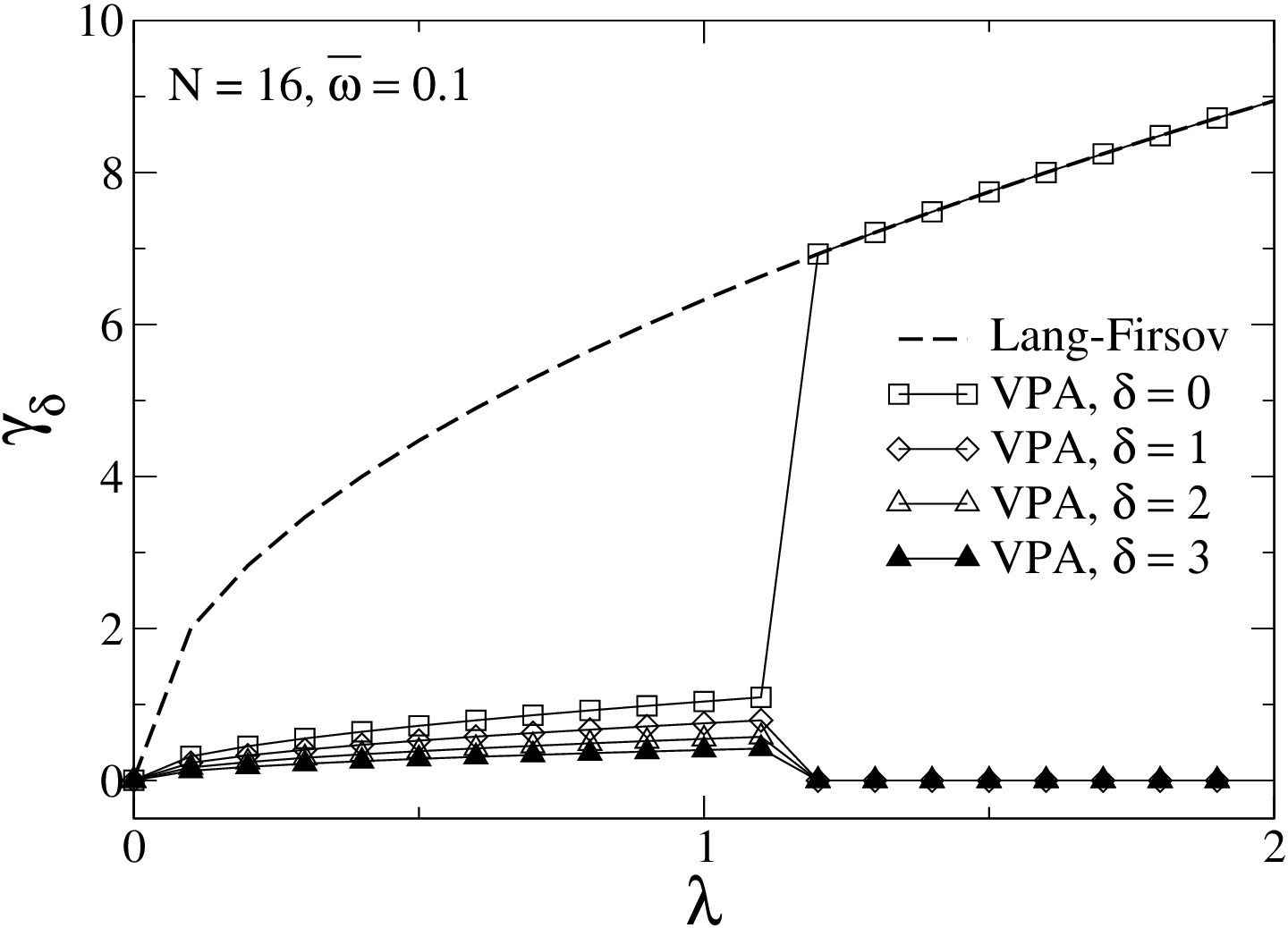}
  \includegraphics[width=0.45\textwidth]{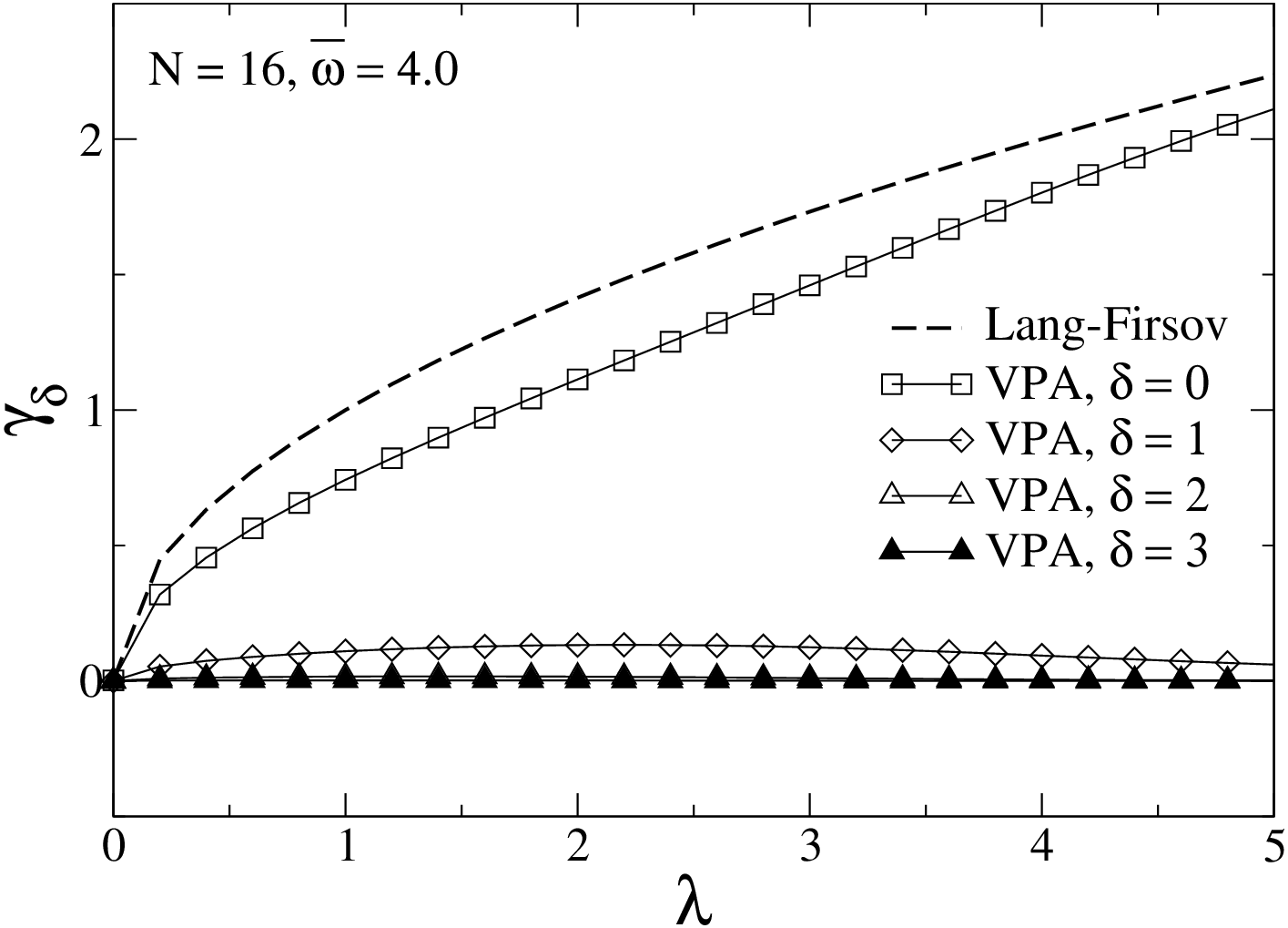}
\caption{\label{fig:gammas}
  Polaron-size parameter $\gamma_\delta$ as a function of the electron-phonon
  coupling $\lambda$ for various distances $\delta$.  The parameter $\gamma$
  of the standard Lang-Firsov transformation (see
  Sec.~\ref{sec:lang-firs-transf}) is also shown.}
\end{figure}

In order to test the validity of VPA we calculated the total energy
[Eq.~(\ref{eq:evs})] and the quasiparticle weight [Eq.~(\ref{eq:z0})] on a
cluster of four sites for various phonon frequencies $\om$ and compared the
results with those of Marsiglio obtained by Lanczos
diagonalization.\cite{Mars95} The comparison is depicted in
Fig.~\ref{fig:E0z0vpa}.  The values of $\omb$ have been chosen to lie in the
nonadiabatic regime $\omb\geq1$ where the zero-phonon approximation of the
VPA is sensible.  The overall agreement is strikingly good.  Minor deviations
from the exact results increase with decreasing phonon frequency.  For the
smallest frequency shown, $\omb=1.0$, the curve for the HLF approximation is
also depicted. It reveals that VPA represents a significant improvement over
the HLF approximation, underlining the importance of the extended polaron
cloud.

The comparison with exact results obtained with Lanczos was
restricted to small clusters with $N=4$ in order to achieve convergence
with respect to the number of phonon states included in the calculation
(see Sec.~\ref{sec:holstein}). To further scrutinize the accuracy of the
VPA we also compare the results of the latter for the total energy
with the variational global-local method which has been shown to give
reliable results over a large range of parameters.\cite{RoBrLi99II} We
chose $N=32$ for which finite-size effects are already very small (see
Sec.~\ref{sec:resQMC}). Moreover, following Romero \etal,\cite{RoBrLi99II}
in Fig.~\ref{fig:E0vpa_GL} we plot $E/\om$ over $g$ with $g=\sqrt{\lambda
W/(2\om)}$. Similar to the case $N=4$ shown in Fig.~\ref{fig:E0z0vpa} we
find a very good agreement for large values of $\omb$ over the whole range of
electron-phonon coupling, whereas for smaller $\omb$ the VPA results begin
to bend away from the correct curve and collapse to the strong-coupling,
atomic-limit result for large $g$. We would like to point out that the
maximum electron-phonon coupling strength in Fig.~\ref{fig:E0vpa_GL}
corresponds to $\lambda\approx 40$ (for $\omb=4.0$), in contrast to
Fig.~\ref{fig:E0z0vpa} where $\lambda\leq2$.
Figures~\ref{fig:E0z0vpa} and~\ref{fig:E0vpa_GL} reveal that in the
nonadiabatic regime $\omb\gg1$ VPA yields a very good agreement with the
exact data and the Global-Local method even in the intermediate and
strong-coupling regime. This behavior can easily be understood considering
the assumptions of the VPA. The zero-phonon approximation becomes exact in
the nonadiabatic limit $\omb\rightarrow\infty$, where the energies of phonon
excitations are too high to have an effect on the ground state.
Finally, we would like to mention the possibility of comparing the
VPA with the QMC results presented in the following section. This has been done
for a variety of parameters, but we have found that it is difficult to
distinguish between deviations due to the shortcomings of the
VPA and due to temperature effects in the QMC results. Consequently, we
have decided to confront the VPA with another approved ground-state method,
namely, the global-local method, which gives a much clearer picture.

In Fig.~\ref{fig:gammas} we show results for the variational displacement
fields $\gamma_\delta$, which give us a measure for the size of the polaron.
For $\omb=0.1$ we see an abrupt crossover from a large to a small polaron at
$\lambda\approx 1.2$. For smaller values of the coupling, the electron
induces lattice distortions at neighboring sites even at a distance of more
than three lattice constants. Above $\lambda\approx1.2$ we have a mobile
small polaron extending over a single site only. In contrast, for a larger
value of the phonon frequency $\omb=4.0$, there is no crossover and we have a
somewhat extended (large) polaron even for large values of $\lambda$. The
same behavior has been found by Marsiglio\cite{Marsiglio95} who determined
the correlation function $\las n_i x_{i+\delta}\ras$ by Lanczos
diagonalization for a restricted phonon basis. Within VPA we have the
relation $\las n_i x_{i+\delta}\ras= \gamma_\delta$.  The main difference is
that in Marsiglio's results, the crossover to a small polaron for $\omb=0.1$
occurs at a smaller value of the coupling $\lambda\approx 1$. Nevertheless,
the simple VPA reproduces the main characteristics of the transition of the
Holstein polaron as the coupling strength $\lambda$ is increased. Finally
Fig.~\ref{fig:gammas} also shows the result for the parameter $\gamma$ of the
standard Lang-Firsov transformation (see Sec.~\ref{sec:lang-firs-transf}).
For $\omb=0.1$, the curves for $\gamma$ and $\gamma_{\delta=0}$ are identical
above the critical value $\lambda\approx 1.2$.  This is not surprising since,
in this regime, we have a small polaron extending over a single site only,
which is well described by the local Lang-Firsov transformation defined in
Eq.~(\ref{eq:LFop}).  For larger values of the phonon frequency (see
Fig.~\ref{fig:gammas}), $\gamma$ and $\gamma_0$ do not coincide above a
critical value of the coupling, but the difference vanishes as
$\lambda\rightarrow\infty$.  In contrast to the adiabatic regime, the polaron
remains an extended object up to very large values of the coupling, so that
the local ansatz of the Lang-Firsov transformation does then not provide the
correct description for finite values of $\lambda$ (see also
Ref.~\onlinecite{AcCuNoRo01}).

\section{\label{sec:resQMC} Results: quantum Monte Carlo}

As our approach is based on a discretized imaginary time, it is important to
study the convergence of any results with increasing number of time slices,
$L$, which determines the error due to the Suzuki-Trotter approximation of
Eq.~(\ref{eq:suzuki-trotter}). $L$ was chosen such that systematic errors are
smaller than the statistical errors of the results. For all observables
considered here we have found the usual $(\Delta\tau)^2$ dependence of the
Suzuki-Trotter error. Depending on the phonon frequency $\omb$ we have found
values of $\dtau=1/30$ (for $\omb\lesssim 1$) and $\dtau=1/40$ (for $\omb>1$)
to be sufficient even for the most accurate results of this paper. Moreover,
as indicated in Fig.~\ref{fig:E0_var_omega}, error bars for the QMC data
presented are always smaller than the symbols used in the figures and are
therefore not shown. Finally, lines connecting data points obtained with QMC
in Figs.~\ref{fig:E0_var_omega}--\ref{fig:p2_var_beta} are guides to the eye
only.

To test our QMC algorithm we have performed several comparisons with other
methods. First, we have checked that the QMC reproduces the exact results
obtained with Lanczos on a four-site cluster. Apart from temperature effects,
an excellent agreement has been found for several different values of the phonon
frequency. Second, as the QMC results are all for finite temperature, we have also
compared them with an exact solution for the two-site system, which is valid
for arbitrary temperature. We have found a perfect agreement over the whole
range of values for $\beta$, $\omb$, and $\lambda$, and can therefore exclude
the possibility of any systematic errors.

\subsection{Kinetic energy}

\begin{figure}
  \includegraphics[width=0.45\textwidth]{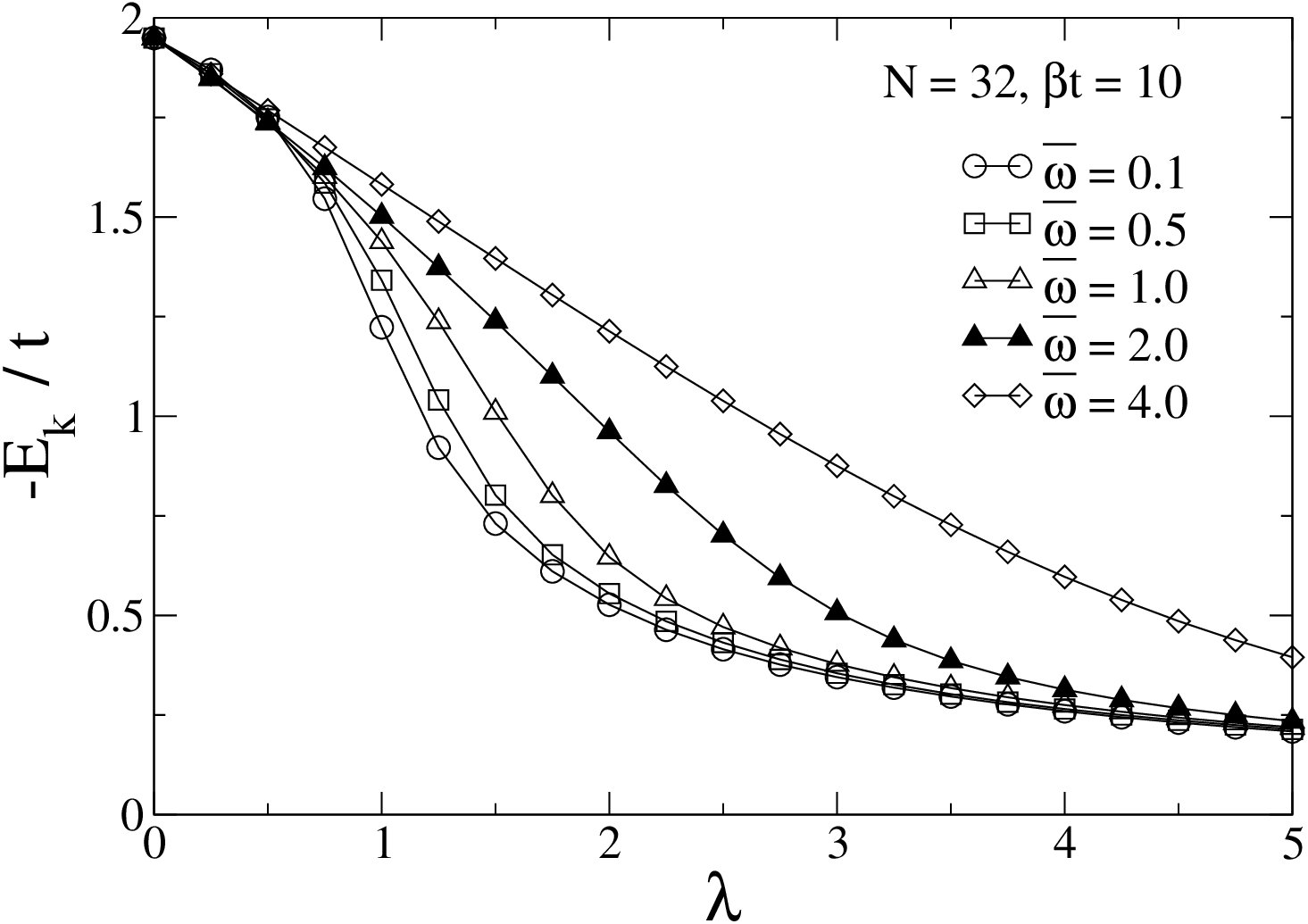}
\caption{\label{fig:Ek_var_w}
  Negative kinetic energy $E_\text{k}$ as a function of the electron-phonon
  coupling $\lambda$ for various values of the phonon frequency $\omb$.  }
\end{figure}

We begin our discussion of the results with the kinetic energy of the
electron, given by Eq.~(\ref{eq:Ekin}), which has previously been calculated
by several authors.\cite{dRLa82,dRLa83,WeRoFe96,Ko97,dMeRa97,RoBrLi99II,JeZhWh99}
In Fig.~\ref{fig:Ek_var_w} we show results for $E_\text{k}$ on a 32-site
cluster, with $\beta t=10$ and for several values of the phonon frequency.
While for small values of $\omb$ there is a rapid decrease of the absolute
value of the kinetic energy in the vicinity of $\lambda=1$, the cross over
becomes smoother as $\omb$ increases. This agrees with the findings of
previous studies and resembles closely the behavior of the total energy
discussed above. For large values of $\lambda$ and $\omb\lesssim1$ we find
$E_\text{k}\sim\lambda^{-1}$ as predicted by small-polaron
theory.\cite{AlMo94} This contrasts strongly with the behavior of the
quasiparticle weight $z_0$ [see Fig.~\ref{fig:E0z0vpa}(b)] which decreases
much faster and is exponentially suppressed in the small-polaron
regime.\cite{JeWh98} As pointed out by Fehske \etal,\cite{FeLoWe00} for the
case of the Holstein model, the quasiparticle weight is exactly the inverse
of the ratio $m_\text{eff}/m$ where $m_\text{eff}$ and $m$ denote the
effective and free mass of the electron, respectively. Hence, in the
small-polaron regime, the effective mass increases exponentially, while the
kinetic energy still has a finite value. We ascribe this behavior to the
undirected motion of the electron inside the phonon cloud, which gives rise
to a nonzero kinetic energy even for large values of $\lambda$. However,
since the polaron bandwidth is exponentially narrowed with increasing
$\lambda$, the polaron is almost localized.

\begin{figure}
  \includegraphics[width=0.45\textwidth]{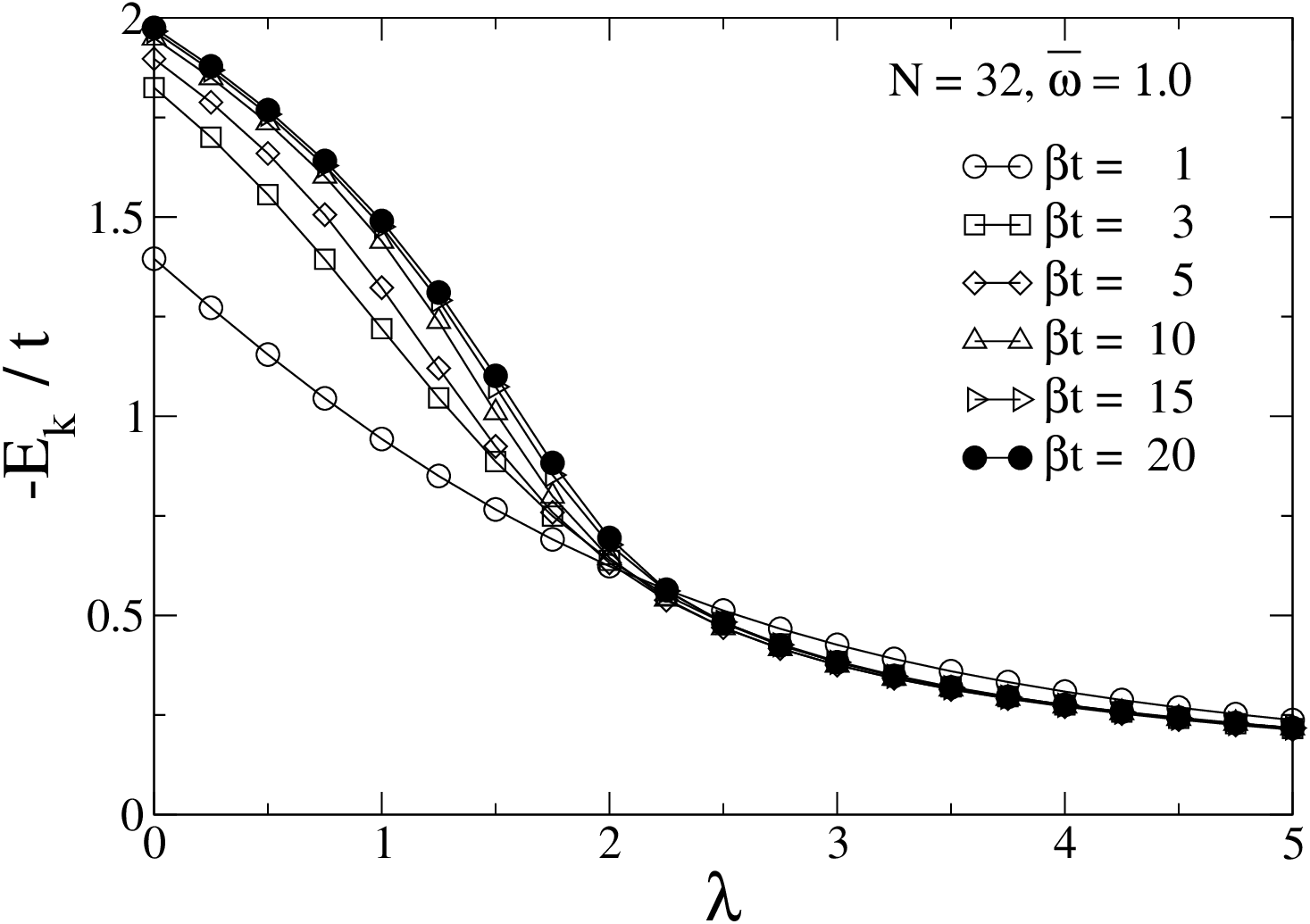}
\caption{\label{fig:Ek_N32_various_beta}
  Negative kinetic energy as a function of the electron-phonon coupling
  $\lambda$ for various values of the inverse temperature $\beta$ for various
  values of the phonon frequency $\omb$.  }
\end{figure}
\begin{figure}
  \includegraphics[width=0.45\textwidth]{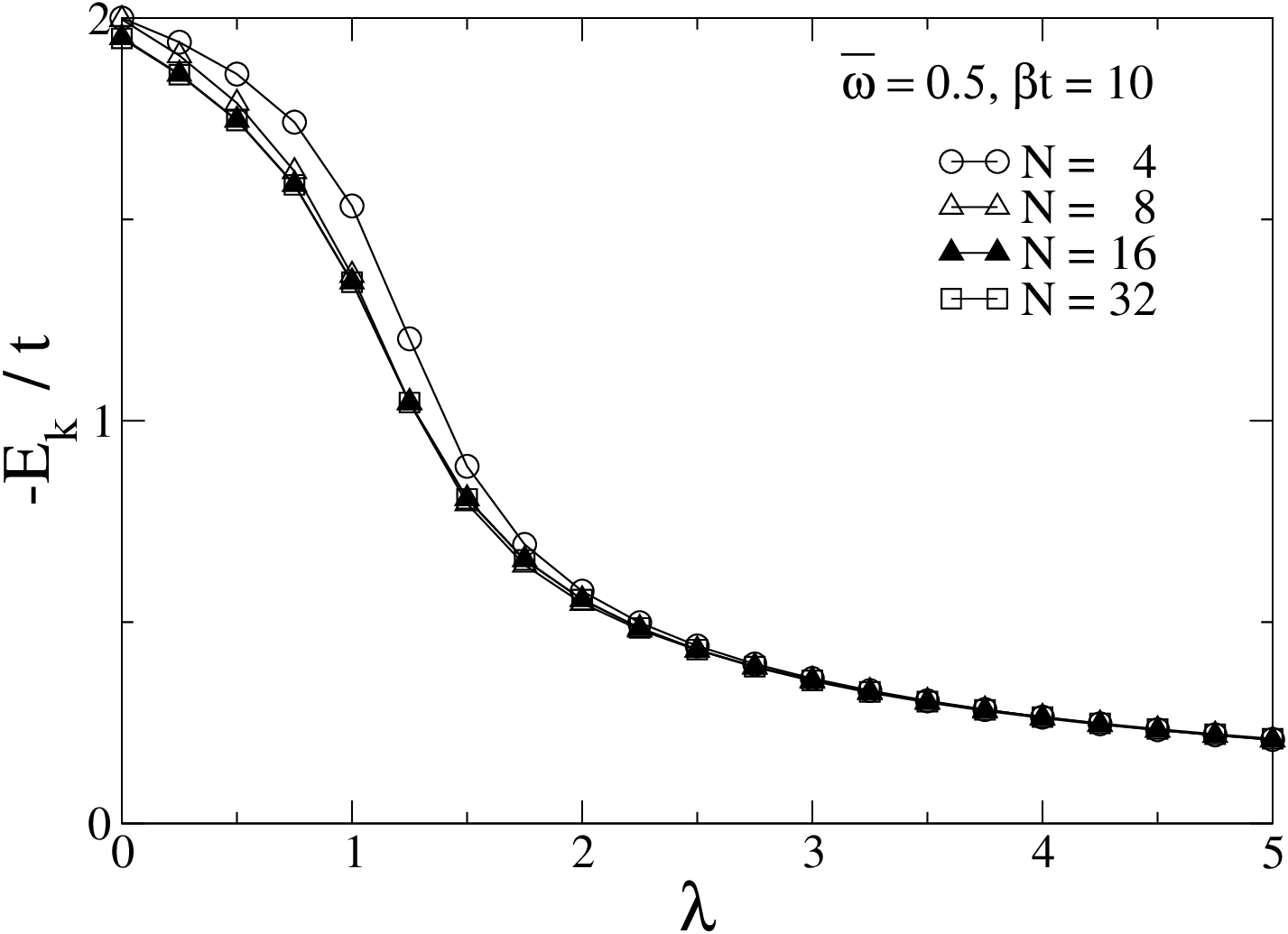}
\caption{\label{fig:Ek_various_N}
  Negative kinetic energy as a function of the electron-phonon coupling
  $\lambda$ for different numbers of lattice sites $N$.}
\end{figure}
\begin{figure}
  \includegraphics[width=0.45\textwidth]{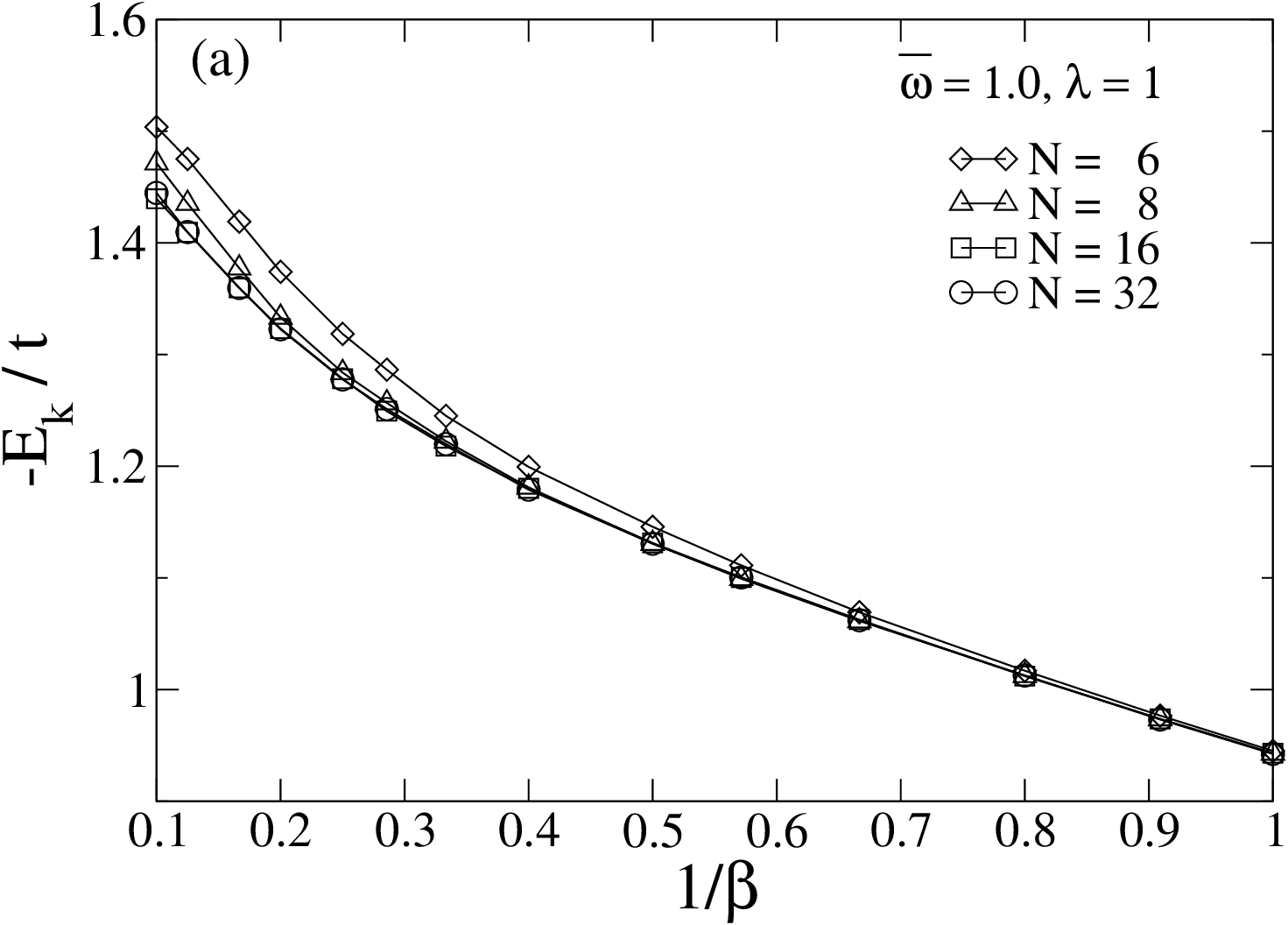}
  \includegraphics[width=0.45\textwidth]{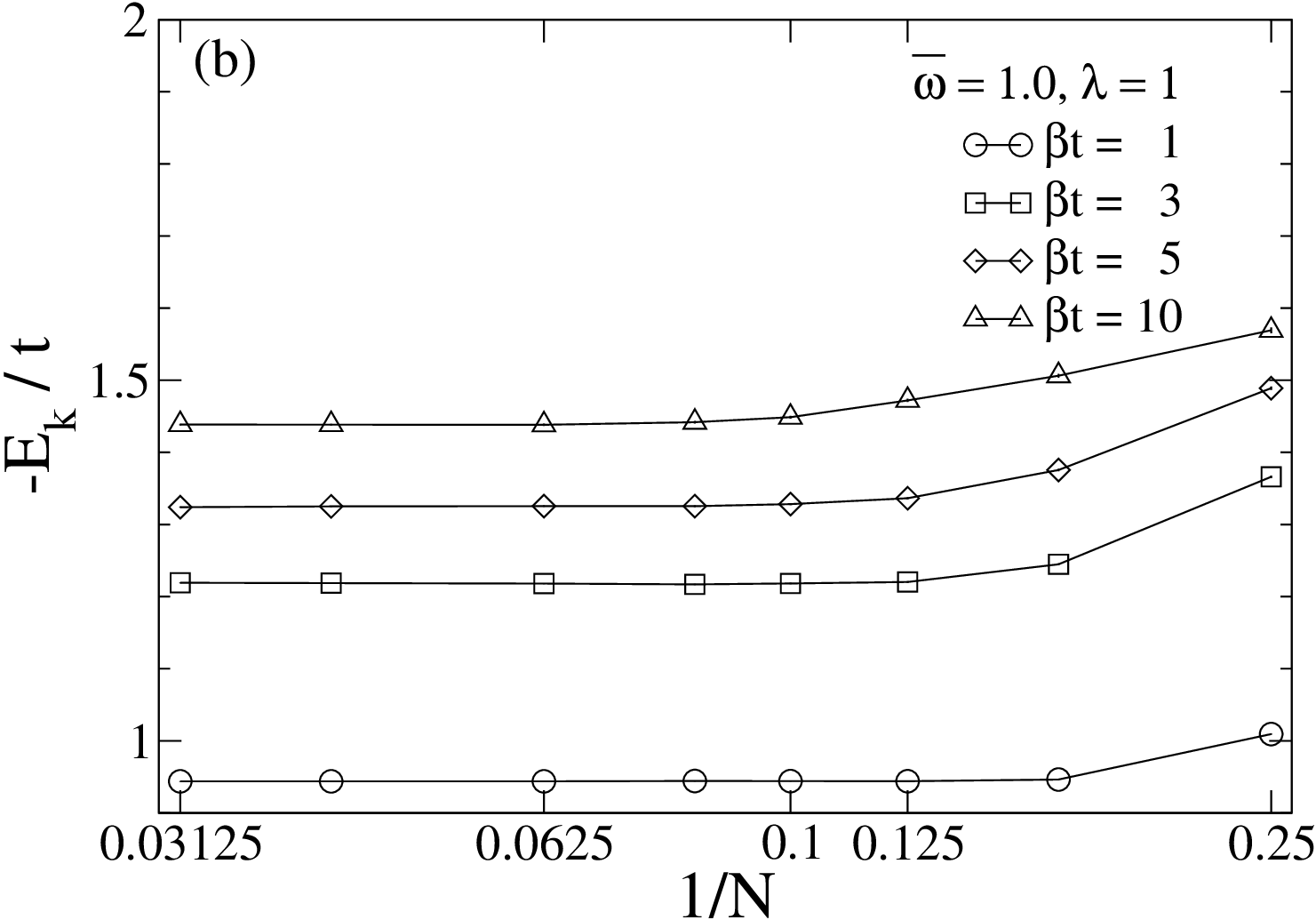}
\caption{\label{fig:Ek_size}
  Negative kinetic energy (a) as a function of the inverse temperature
  $\beta$ and (b) as a function of the inverse of the number of lattice sites
  $N$.}
\end{figure}

To study the influence of temperature we have calculated the kinetic energy
for a system of 32 sites, with $\omb=1.0$ and for various values of $\beta t$
(see Fig.~\ref{fig:Ek_N32_various_beta}). As $\beta t$ increases,
$|E_\text{k}|$ increases for $\lambda\lesssim2$. However temperature effects
are obviously very small in the strong-coupling regime. For $\beta t=1$,
$|E_\text{k}|$ decays in a smooth way as $\lambda$ is increased, while for
lower temperatures we find the typical rather abrupt crossover near
$\lambda=1$, as in Fig.~\ref{fig:Ek_var_w}. De Raedt and
Lagendijk\cite{dRLa83} have calculated the kinetic energy for the same set of
parameters using their QMC method. However, the lowest temperature they could
reach was $\beta t=5$ which, according to Fig.~\ref{fig:Ek_N32_various_beta},
is still quite different from the ground-state result. Moreover, their
calculations did not include dynamical effects of the phonon degrees of
freedom. As a consequence, for $\beta t=1$, they do not obtain the correct
behavior of the kinetic energy as a function of $\lambda$. Finally, Romero
\etal\cite{RoBrLi99II} and Jeckelmann and White\cite{JeZhWh99} calculated the
kinetic energy for $T=0$ on a 32-site cluster and for an infinite system,
respectively. Their results are in a good agreement with our findings, although
small deviations due to temperature and finite-size effects are
visible. Nevertheless, we can conclude from Fig.~\ref{fig:Ek_N32_various_beta}
that a value of $\beta t=10$ should be sufficient to obtain results which are
representative of the ground state.

We now turn our attention to finite-size effects.  In
Fig.~\ref{fig:Ek_various_N} we show the kinetic energy for $\omb=0.5$, $\beta
t=10$, and for various number of lattice sites.  For $N\geq16$ the results for
$E_\text{k}$ are well converged over the whole range of $\lambda$ and
finite-size effects are very small. This agrees with the findings of other
authors.\cite{dRLa83,Marsiglio93,Marsiglio95} Figure~\ref{fig:Ek_size} shows
the kinetic energy as a function of temperature, for $\omb=1.0$ and various
numbers of lattice sites $N$. Moreover we chose $\lambda=1$, as the influence
of the system size is largest in the cross over regime. Figure~\ref{fig:Ek_size}(a) clearly
demonstrates that finite-size effects are most pronounced at low
temperatures, while they are completely smeared out at higher temperatures,
since high-temperature properties are determined by integral quantities, such
as energy moments $\langle E^\nu\rangle$, which have a small size dependence,
while low-temperature features are governed by energetically low-lying
eigenvectors. To further illustrate the influence of the system size, we plot
in Fig.~\ref{fig:Ek_size}(b) the negative kinetic energy as a function of
$1/N$ again for $\omb=\lambda=1$ and for various values of $\beta$. As
before, error bars are smaller than the symbol size, but due to the very high
accuracy of the data, the systematic errors due to the finite number of
Trotter slices $L$ are comparable to the statistical errors. The results show
that very good convergence with
respect to the number of lattice sites is achieved for rather small $N$. In
fact, for the highest temperature shown ($\beta t=1$), the line connecting
the data points becomes vertical already at $N=8$, while for $\beta t=10$
convergence is reached for $N=16$. Hence, if we consider these findings in
the context of the usual finite-size scaling analysis where one plots the
data as a function of a suitably chosen power of $1/N$ and extrapolates to
the infinite system (\ie, $1/N\rightarrow0$), we have here the special case of
a linear dependence with zero slope at large enough $N$. Thus, in contrast to
the half-filled Holstein model of spinless fermion, for which a finite-size
analysis has been performed by two groups,\cite{KeHaMu96,WeFe98} we merely
find that the results converge within the accuracy of our calculations
already for rather small systems.

\subsection{Total energy}

\begin{figure}
  \includegraphics[width=0.45\textwidth]{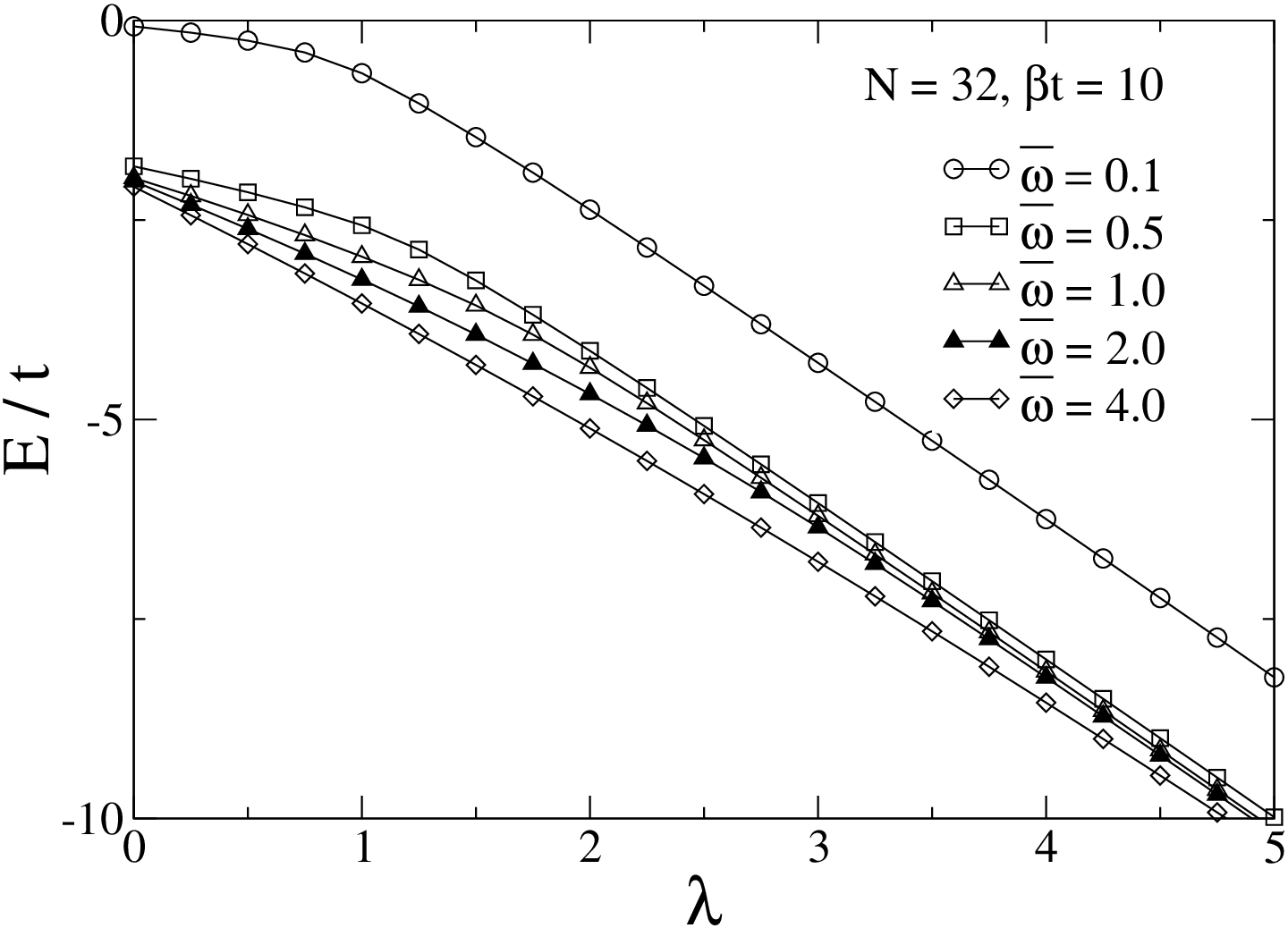}
\caption{\label{fig:E0_var_omega}
  QMC results for the total energy $E$ as a function of the electron-phonon
  coupling $\lambda$ for various values of the phonon frequency $\omb$. Here
  and in subsequent figures lines are guides to the eye, and errorbars for
  the QMC data are smaller than the symbols shown.}
\end{figure}

Next we consider the total energy $E$, given by
Eq.~(\ref{eg:E0}). In Fig.~\ref{fig:E0_var_omega} we present the total energy
for a cluster of 32 sites and various values of the phonon frequency. Finite
temperature effects increase as we approach the low-frequency regime, and for
$\omb=0.1$ we clearly see a strong deviation from the ground-state result
$E=-2t$ for $\lambda=0$. The frequency-dependence of the temperature effects
can easily be understood if we consider the exact result for
the kinetic energy of $N$ independent harmonic oscillators
\begin{equation}\label{eq:harm-osc}
  E_\text{k,ph} = \frac{\om}{2}\sum_i\las
  p_i^2\ras=\frac{N\om}{2}\left(\frac{1}{2}+
  \frac{1}{e^{\beta\om}-1}\right)\,,
\end{equation}
which is identical to the second term in Eq.~(\ref{eg:E0}). For low
temperatures we have $\las p^2\ras\approx0.5+e^{-\beta\om}$, with a
correction to the ground-state value of $0.5$ that increases with decreasing
$\om$.  These temperature effects on $E$ due to the oscillator energy do
not depend on $\lambda$ [see Eq.~(\ref{eq:harm-osc})] and therefore shift the
total energy curve by the same amount for all values of the coupling.
A comparison with the discussion of the kinetic energy reveals that
temperature effects are much smaller for other observables due to the absence
of the strongly temperature-dependent phonon energy terms $P_\text{p}$ and
$E'_\text{ph}$ [see Eqs.~(\ref{eq:holstein}) and~(\ref{eg:E0})].

The dependence on $\omb$ agrees well with existing
work.\cite{dRLa82,dRLa83,RaTh92,AlKaRa94,Mars95,Marsiglio95,CaCiGr98,Ko97,Ko98,RoBrLi99I,RoBrLi99II,FeLoWe00}
It is known\cite{Marsiglio95} that at zero temperature and for small values
of the phonon frequency, $\omb\lesssim 0.5$, the total energy displays a
rather sharp transition around $\lambda\approx1$, where the cross over from a
large to a small polaron occurs. In ED studies of small
clusters\cite{Marsiglio95} a kink in $E$ has been observed, which is smeared
out in the finite-temperature QMC results.  Nevertheless, we observe the same
rounding of the energy curve with increasing $\omb$
(Ref.~\onlinecite{Marsiglio95}). As discussed by Marsiglio\cite{Marsiglio95}
the kink in the total energy is merely a finite-size effect. As the system
size increases the discontinuity disappears, in accordance with the fact that
the ground state of the Holstein polaron is an analytic function of the
coupling parameter $\lambda$ (Ref.~\onlinecite{Loe88}).

Finally, it is interesting to note that in contrast to the kinetic
energy $-E_\text{k}$, which shows a sharp decrease near $\lambda=1$ in the
adiabatic regime (see, \eg, Fig.~\ref{fig:Ek_var_w}), the total energy does not
change significantly. As discussed for the two-dimensional case
by Kornilovitch,\cite{Ko97} this can be explained as follows. For small
ratios $\om/t$, the phonon energy associated with the term $P$ of
Hamiltonian~(\ref{eq:holstein}) is small and the system is governed by the
balance of the electronic kinetic energy and the energy due to the
electron-phonon coupling. In the transformed model, the latter is given by
$E_\text{p}$ as defined by Eq.~(\ref{eq:Ep}). When the ratio of the two
energies approaches unity (equivalent to $\lambda=1$), it becomes
energetically favorable for the electron to localize (losing kinetic
energy) and increase its potential energy. This leads to finite
displacements of the oscillators in the vicinity of the electron and
increases the potential energy of the phonons. Hence, near $\lambda=1$ the
energy of the system is redistributed from kinetic to potential energy so
that $E$ remains almost unchanged. This is exactly what we see in
Fig.~\ref{fig:Ek_var_w}.

\subsection{Momentum distribution and oscillator momenta}

\begin{figure}
  \includegraphics[width=0.45\textwidth]{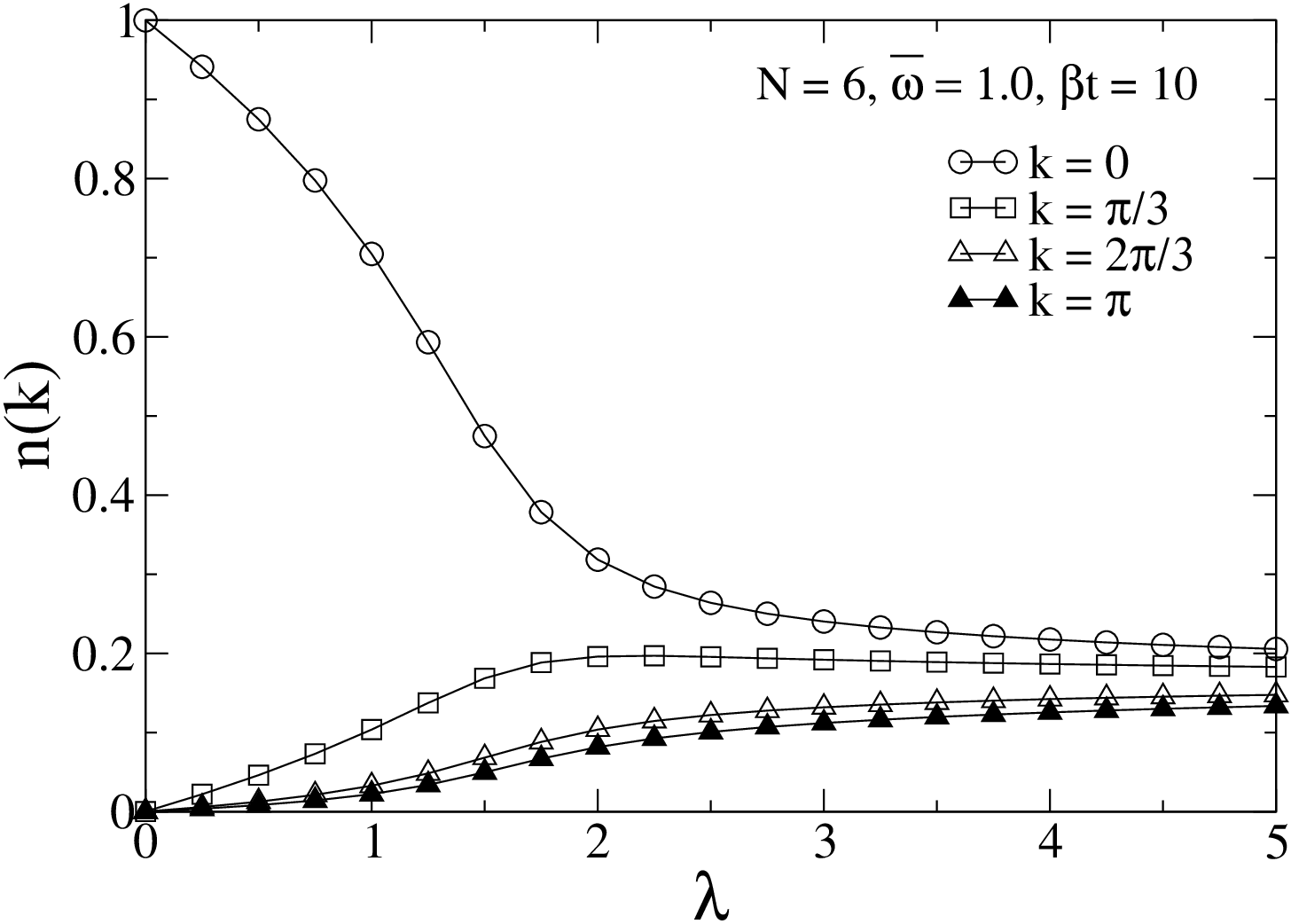}
\caption{\label{fig:nk_var_k}
  Momentum distribution $n(k)$ as a function of $\lambda$ for
  various wave vectors $k$.}
\end{figure}

Following Zhang \etal\cite{ZhJeWh99} we also calculated the momentum
distribution $n(k)$, given by Eq.~(\ref{eq:nk}), for different wave vectors
$k$ (Fig.~\ref{fig:nk_var_k}). To compare with their DMRG\footnote{The
  coupling parameter $g$ used in Ref.~\onlinecite{ZhJeWh99} is related to
  ours by $\lambda=2g^2/(\om t)$.} results we chose the same parameters $N=6$
and $\omb=1.0$. Moreover, we took $\beta t=10$ since the calculations of Zhang
\etal were for the ground state. For $\lambda=0$ the ground state has
momentum $k=0$, so we have $n(0)=1$ and $n(k\neq0)=0$. With increasing
coupling $n(0)$ decreases in a way similar to the kinetic energy (cf.
Fig.~\ref{fig:Ek_var_w}), while $n(k)$ for $k\neq0$ increases. In the
strong-coupling limit $\lambda\rightarrow\infty$, $n(k)$ approaches the value
$1/N=1/6$ for all $k$. This is a simple consequence of the localization of
the electron for $\lambda=\infty$. Although the curve for $k=0$ looks very
similar to the results of Zhang \etal we find a slightly stronger decrease of
$n(0)$ in the intermediate coupling regime. This deviation is no temperature
effect of our QMC method but probably originates from the fact that Zhang
\etal obtained their results for $n(0)$ by integrating over an approximate
spectral function.

\begin{figure}
  \includegraphics[width=0.45\textwidth]{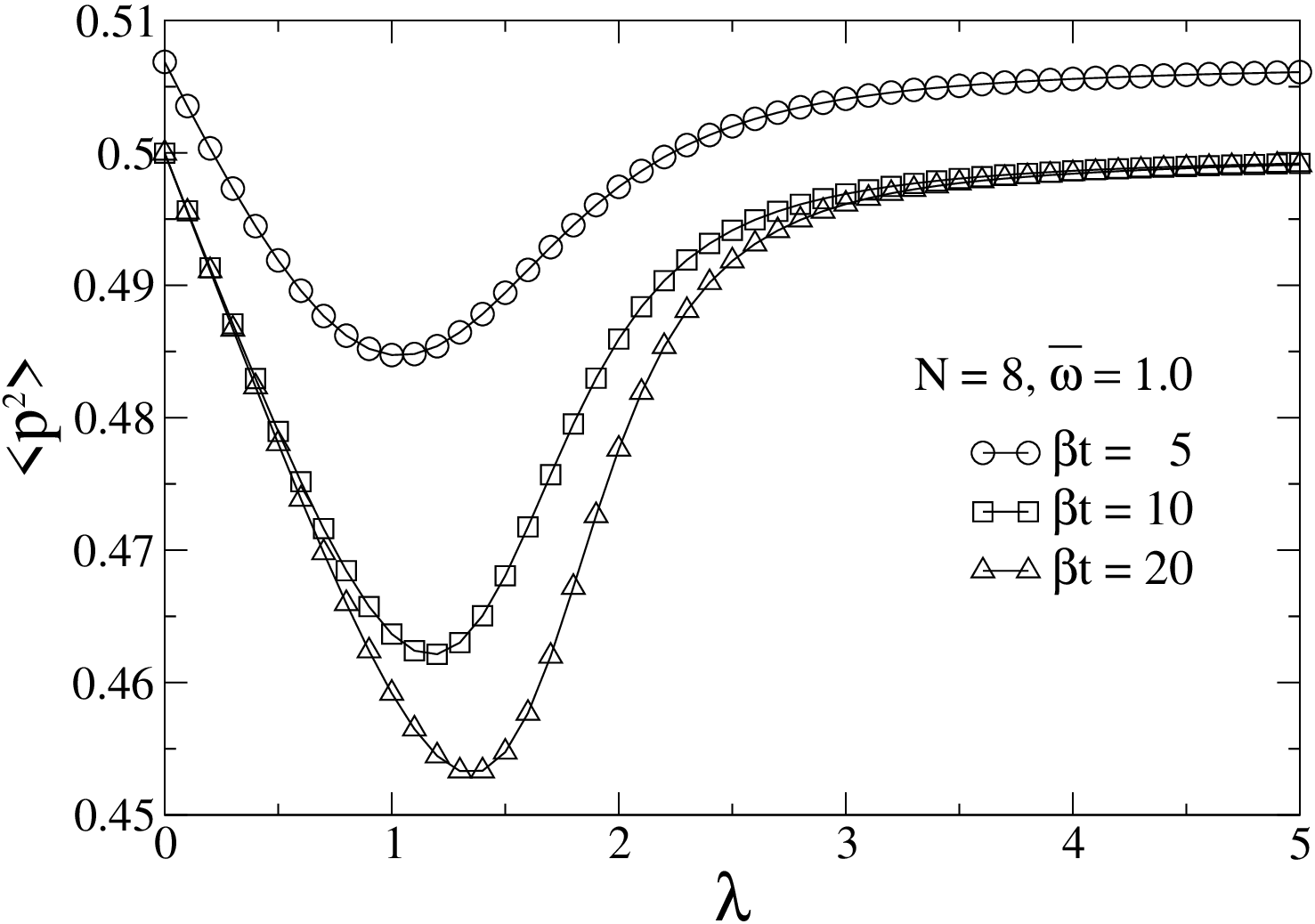}
\caption{\label{fig:p2_var_beta}
  Mean square of the phonon momentum $p$ as a function of the electron-phonon
  coupling $\lambda$.}
\end{figure}

In Sec.~\ref{sec:reweighting} we mentioned that, within the Lang-Firsov
approach, the phonon degrees of freedom only show a weak dependence on the
electron-phonon coupling, in contrast to the standard approach, where the
average oscillator coordinate $\las x\ras$ increases strongly with $\lambda$
due to the displacement in the presence of an electron. The weak dependence
of the vibrational energy of the local oscillator, which is proportional to
$\la p^2\ra$, on $\lambda$ is shown in Fig.~\ref{fig:p2_var_beta}. For
$\lambda=0$ we have the result $\las p^2\ras=0.5 + [\exp(\beta\om)-1]^{-1}$
[see Eq.~(\ref{eq:harm-osc})] for a free oscillator. In the intermediate
coupling regime, $\las p^2\ras$ takes on a minimum, corresponding to a
reduction of merely $4\%$ and approaches the value for $\lambda=0$ again in
the strong-coupling limit. As the Lang-Firsov transformation does not affect
the phonon momenta $p$ (see Sec.~\ref{sec:lang-firs-transf}), the result for
$\las p^2\ras$ as a function of $\lambda$ is the same in the untransformed
Holstein model. However, the significant advantage of the proposed method is
that the phonon momenta are sampled instead of the coordinates $x$. Thus the
probability distribution associated with the degrees of freedom to be sampled
has only a small variance compared to the standard method, which makes the
simulations much more effective. The dependence of $\las p^2 \ras$ on the
coupling strength $\lambda$ and the temperature has first been studied by
Ranninger and Thibblin\cite{RaTh92} for the two-site polaron problem.  For
such a small system, the minimum of $\las p^2\ras$ is even more pronounced,
while for larger systems the average effect of the electron on a local
oscillator is more and more washed out.  Ranninger and Thibblin\cite{RaTh92}
ascribed the deviation of the vibrational energy from the free-oscillator
result to anharmonic effects, which are visible only at low enough
temperatures. This can clearly be seen in Fig.~\ref{fig:p2_var_beta}, where
the minimum of $\las p^2\ras$ becomes less pronounced and is shifted to
smaller values of $\lambda$ as the temperature increases.

\subsection{Performance}

We conclude this section with a discussion of the performance of the QMC
approach. From the results presented above it is obvious that the method
enables us to study a very wide range of parameters. Hence, for example, we
have performed calculations for $0.1\leq\omb\leq4.0$ (see
Fig.~\ref{fig:Ek_var_w}). Simulations in the adiabatic regime would be
extremely difficult within the standard approach, since the autocorrelation
times grow as $(\om\dtau)^{-2}$. However, in materials such as the
manganites, the frequencies of the relevant phonon modes are known to be
small ($\omb\lesssim 0.5$, see, \eg, Ref.~\onlinecite{David_AiP}) so that our
method could represent an important step forward towards the simulation of
electron-phonon models with realistic parameters. Also, we are able to reach
very low temperatures $\beta t\leq20$ and clusters large enough to avoid
finite-size effects with modest computational effort. Another key advantage
is that the method becomes more and more efficient as the coupling strength
$\lambda$ increases, which is due to the use of the Lang-Firsov
transformation. In our results we find that statistical errors of expectation
values of phonon operators are larger than, \eg, the errors of the kinetic
energy. Finally, the errors increase slightly as we approach the adiabatic
and/or low-temperature regime $\omb\rightarrow0$ and $\beta
t\rightarrow\infty$, respectively.

To demonstrate the efficiency of our method we give some figures for the CPU
time of the simulations. A typical QMC run for 32 lattice sites, $\beta t=5$,
$\omb=1.0$, and $\lambda\approx 1$ (\ie, near the small-polaron crossover) only
takes 5 min of CPU time on a 650 MHz Pentium III PC. For such a run
relative errors of, for example, the kinetic
energy are less than 1.0\%. Away from the crossover point, the same accuracy
can be obtained within a few seconds. For $\beta t=10$, the temperature used
in most of the calculations presented in this paper, a QMC run with $\lambda$
near the crossover value and with similar statistical errors as mentioned above
takes about 80 min of CPU time. Hence, although not as efficient as the
specialized one-electron methods,\cite{dRLa82,dRLa83,dRLa84,Ko97,Ko98,Ko99}
our approach significantly reduces the numerical effort compared to
previous methods which were often run on supercomputers and did not reach the
parameters (low temperature and small phonon frequency) and accuracy of our
simulations.

\section{\label{sec:summary}Conclusions}

We have presented a simple variational approach to the Holstein model, which
incorporates an extended Lang-Firsov transformation. This approach is easily
applicable to infinite systems and represents a marked improvement over the
standard small-polaron approximation, which is only useful in the
nonadiabatic, strong-coupling regime.

More importantly, we have introduced an exact QMC method for the Holstein
model, which is based on the standard Lang-Firsov transformation of the
Hamiltonian. The phonon momenta are represented in terms of principal
components, which enables us to sample completely uncorrelated
configurations, while the electronic degrees of freedom are taken into
account exactly by use of a reweighting method for calculating observables.
Thereby, we avoid the numerically expensive evaluation of the electronic
weights in the updating process. The present approach can be applied for a wide
range of parameters with relatively small computational effort. In
particular, efficient simulations can be performed in the adiabatic regime
$\omb<1$, which is of special interest in connection with materials such as
the manganites. In the one-dimensional case considered here, a sign problem
resulting from the Lang-Firsov transformation on small systems has been found
to have only a small effect on the statistics.  Tests have been presented in
the one-electron case and reveal that the method reproduces Lanczos
diagonalization results in the regime where the latter are applicable, namely,
for very small systems, small to moderate electron-phonon coupling and for
sufficiently large phonon frequency. Moreover, a satisfactory agreement with
other methods has been found. Owing to the exact treatment of the electronic
degrees of freedom and the sampling of the phonons, the method is free of any
autocorrelations. The use of the Lang-Firsov transformation, which is
essential for the applicability of the reweighting method, substantially
improves the statistics, allowing for very accurate results.

Despite the great computational efficiency of our method compared to the
standard approach, even faster methods exist. For example, the QMC
simulations of de Raedt and Lagendijk\cite{dRLa82,dRLa83,dRLa84} and
Kornilovitch\cite{Ko97,Ko98,Ko99} seem to be numerically faster due to the
analytic integration over the phonon degrees of freedom which significantly
reduces statistical errors. However, both methods are restricted in their
applicability as discussed in Sec.~\ref{sec:introduction}. In particular, an
extension to many-electron systems seems impossible, since simulations will
be restricted by a severe minus sign problem similar to other world-line
methods. In contrast, the method presented here is not restricted
to the single-electron limit in principle, although some modifications will
be necessary.  As pointed out in previous sections, most of the ideas
proposed here, such as the use of the transformed model, the reweighting
method, and the PC representation, remain unchanged if we consider more than
one electron. The required modifications concern mainly the fermionic weight
$\wf$ [Eq.~(\ref{eq:omega})]. There is a Hubbard-like interaction term coming
from the Lang-Firsov transformation (see Sec.~\ref{sec:lang-firs-transf}),
and the one-electron basis states used here (Eq.~(\ref{eq:oneelbasis}) have
to be replaced by the corresponding set of many-electron states. Since the
number of such basis states, and thereby the linear dimension of the matrices
$\Omega$, $\kappa$, and $D$ (see Sec.~\ref{sec:QMCI}), grows exponentially
with the system size, an exact treatment of the fermion degrees of freedom
will become increasingly difficult. Consequently, a more refined approach
based on, \eg, the use of determinant methods will be required.  For the
bipolaron problem of two electrons with opposite spin, on which work is
currently in progress, the computational effort can be significantly reduced
by exploiting the conservation of the total quasimomentum (see, \eg,
Ref.~\onlinecite{KoPi97}), leading to a computer time that grows with the
cube of the system size. The more general case of, \eg, a quarter-filled band
corresponding to the colossal magnetoresistance regime of the manganites,
requires further consideration, and the effect of the sign problem remains to
be investigated. Moreover, the performance of such an approach has to be
compared to existing many-electron QMC methods for the Holstein model.
Finally, the method can be generalized to more complicated
models including, \eg, a coupling of the electrons to local spins as in the
Kondo or double-exchange model for the manganites.

\begin{acknowledgments}
  
  This work was partially supported by the Austrian Science Fund (FWF),
  project No.~P15834. M.H. is grateful to the Austrian Academy of Sciences for financial support. We would like to
  acknowledge helpful discussions with Markus Aichhorn, Holger Fehske,
  Winfried Koller, and Alexander Pr\"ull. We would also like to thank Frank
  Marsiglio and Aldo Romero for providing us with some of the data presented
  in this paper.

\end{acknowledgments}



\end{document}